# On the Determination of the Solar Rotation Elements *i*, Ω and Period using Sunspot Observations by Ruđer Bošković in 1777

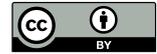

**Mirko Husak[1], Roman Brajša[2], Dragan Špoljarić[3]**
[1] Trakošćanska 20, 42000 Varaždin, Croatia, https://orcid.org/0000-0002-2814-5119
[2] Geodetski fakultet Sveučilišta u Zagrebu, Opservatorij Hvar, Fra Andrije Kačića Miošića 26, 10000 Zagreb, Croatia
[3] Geodetski fakultet Sveučilišta u Zagrebu, Fra Andrije Kačića Miošića 26, 10000 Zagreb, Croatia

**Abstract**
In September 1777, Ruđer Bošković observed sunspots for six days. Based on these measurements, he used his own methods to calculate the elements of the Sun's rotation, the longitude of the node, the inclination of the solar equator and the period. He published a description of the methods, the method of observation and detailed instructions for calculations in the second chapter of the fifth part of the *Opera* in 1785. In this paper, Bošković's original calculations and repeated calculations by his procedure are published. By analysing the input quantities, procedures, and results, the input quantities of the error, and the calculation results are discussed. The reproduction of Bošković's calculations is successfully reproduced and we obtained very similar results. The conclusion proposes a relationship of Bošković's research with modern astronomy.

## 1. Introduction

Ruđer Josip Bošković (1711 - 1787) was a Croatian Jesuit priest with a broad interest in various scientific fields (**Dadić, 1998; James, 2004; MacDonnell, 2014**). He made important contributions to mathematics, astronomy, physics, geodesy, cartography, archeology, civil engineering, and philosophy. Moreover, he was a successful poet and diplomat, and invented/improved some optical and astronomical instruments.

Astronomy was, however, one of his main research fields (**Kopal, 1961; Špoljarić and Kren, 2016; Špoljarić and Solarić, 2016**). R. Bošković investigated, both theoretically and observationally, many astronomical phenomena: solar rotation and sunspots, transit of the planet Mercury before the Sun, the aberration of stars, eclipses, comet and planetary orbits, among others. He was also a founder of the Brera Observatory in Italy, which confirms a dominant interest of R. Bošković in astronomy.

In his first scientific paper (**Boscovich, 1736**), Bošković introduced and described two methods for the determination of the solar rotation elements: the solar rotation axis position in space and the rotational period. The direction of the solar axis in space is determined by the angle of inclination (*i*) between the ecliptic plane and the equatorial plane of the Sun and by the ecliptic longitude (Ω) of the ascending node of the solar equator, which is the angle, in the ecliptic, between the equinox direction and the direction where the solar equator intersects the ecliptic from the South, i.e. in the sense of rotation (**Stix, 2002**). In his last extensive astronomical work (**Boscovich, 1785**), Bošković described his third method for the determination of the solar rotation elements, and applied the methods to his own sunspot observations performed in Sens, near Paris in 1777.

Solar rotation elements belong to the fundamental astronomical properties of the Sun (**Stix, 2002**). A precise knowledge of the solar rotation axis position in space is important for data reduction of solar observations and transformation of the measured coordinates of objects on the Sun into heliographic ones (**Wöhl, 1978; Stark and Wöhl, 1981; Balthasar et al., 1986; Balthasar et al., 1987; Reinsch, 1999; Stix, 2002**). Solar rotation can be expressed in terms of the rotational period (in days) or in terms of the angular velocity (in degrees per day). The investigation of solar rotation is very important within solar physics. The Sun rotates differentially, which means that the angular velocity is a function of the heliographic latitude, the depth and the time (**Howard, 1984; Schröter, 1985; Beck, 1999**), which is possible since the Sun is composed of plasma and does not rotate as a rigid body. Moreover, research of the solar differential rotation is important, as it is closely related to the solar magnetohydrodynamical dynamo, which,

Corresponding author: Mirko Husak
*mhusak@geof.hr*

according to present concepts, plays a significant role in generating and maintaining solar magnetic activity (**Stix, 2002; Charbonneau, 2020**).

The three main experimental methods to measure solar rotation are the tracer method, the Doppler method and the helioseismological method (**Beck, 1999**). R. Bošković (**Boscovich, 1785**) used the tracer method, which is also the oldest method for the solar rotation determination. The method consists of following the positions of any recognizable objects on the Sun, e.g. sunspots, in time. Then one possibility is to determine the solar angular velocity of rotation by dividing the difference in position with the elapsed time, and this is the procedure used by R. Bošković.

There is a large quantity of observational evidence which shows that the solar rotation changes in time (**Brajša et al., 1997; Brajša et al., 2006; Wöhl et al., 2010**) and that this variation is related to the solar activity cycle (**Jurdana-Šepić et al., 2011; Ruždjak et al., 2017**). For this reason, it is important to discover, collect and reduce as many historical sunspot observations as possible (**Arlt and Vaquero, 2020; Nogales et al., 2020**).

In this present work, we describe the three methods of R. Bošković for the determination of the solar rotation elements. We repeat the original calculation of R. Bošković applied to his own sunspot observations with the aim to fully understand the method and to reproduce his published results.

## 2. Methods of Ruđer Bošković for the determination of the inclination $i$, the longitude of the node $\Omega$, and the period of the solar rotation $T$ using sunspot observations

Ruđer Bošković developed three methods for the determination of the solar rotation elements: the inclination $i$, the longitude of the node $\Omega$, and the period of solar rotation: the graphical method, and two numerical methods using the planar and the spherical trigonometry (**Boscovich, 1785, Préface, №4-9, pages 77-79**). The original formulas and descriptions of figures for his methods are presented in *Opuscule II* (**Boscovich, 1785, №10-19, pages 79-85**). At the end of Tomus V, the table of figures 1. to 9. and the extract (**Boscovich, 1785, Extrait, starting page 444, §.II. Du second Opuscule, №25-№33, pages 456-461**) are presented.

The methods for determining the elements of solar rotation, based on the observation of sunspots, Ruđer Bošković published in 1736 in his first dissertation *De maculis solaribus* (*About sunspots*) in Latin (**Boscovich, 1736**). The formulas for the methods of Ruđer Bošković are presented in the *Préface* (**Boscovich, 1785, Préface, pages 81-84**). The example is performed with logarithmic tables presented in twelve Roman numbered tables *Tab. I.* to *Tab. XII.* (**Boscovich, 1785, pages 166-169**). The methods will be described here with stress on the reproduction of Bošković's example, but not the method's analysis. The methods are described in *Opuscule II* (**Boscovich, 1785, §.II.–§.VII., №27-81, pages 89-118**).

### 2.1. The method for $\Omega$

The method for the determination of the longitude of the node $\Omega$, the intersection of the ecliptic and the solar equator, uses two positions of the same sunspot on the equal latitude. The observation of the same sunspot on the same latitude practically is not reliable, but it can be mathematically simulated using three sunspot positions, the one on the left, and the two on the right of the maximal sunspot latitude (**Boscovich, 1785, §.IV., №45**). The numerical solution of the method is the ratio of the differences of ecliptic latitudes and longitudes.

### 2.2. The method for $i$

The planar trigonometric method for the determination of the inclination of the solar equator regarding the ecliptic uses two sunspot positons with the longitude of the node that is already known. The method uses one planar triangle for inclination determination: the planar graphical construction and the planar trigonometry solution (**Boscovich, 1785, §.V., №53**).

### 2.3. The method for the period

The method for the solar rotation period determination uses two sunspot positions and already known elements: the longitude of the node and the inclination of the solar equator. The method uses spherical trigonometry for determination of the angle between two declination arcs from two sunspot positons to the pole in the equatorial coordinate system. The period of rotation is calculated from the angular velocity (**Boscovich, 1785, §.VI.**).

### 2.4. The method for $i$ and $\Omega$

The method for determining the two elements of solar rotation calculates: the longitude of the node and the inclination of the solar equator regarding ecliptic, using three positions of the same sunspot (**Boscovich, 1785, §.VII.**). The method has three solutions: 1. the graphical method (**Boscovich, 1785, №70**), 2. the planar trigonometry method (**Boscovich, 1785, №69**), and 3. the spherical trigonometry method (**Boscovich, 1785, №76-№79**).

## 3. Results

Bošković published his results in *Opuscule II* (**Boscovich, 1785, Tab. I. to XII., pages 166-169**). In this present work, these are **Table 8** to **Table 19**, and the re-production of his results in this present work are: **Table 22** to **Table 33**. In *3.4. Present work results*, only input and output calculation data are displayed instead of larger tables. Tables in the present work are formatted as much as possible like the original tables using fonts and formatting rules (upright, italic, bold). In the beginning (**Boscovich, 1785, №3**) Bošković described what we need for the reproduction of his results: the sunspot observations, the astronomic almanac (**URL 1**), *Opuscule II* (**Boscovich, 1785**) and logarithmic tables. The formulas are trigonometric. The original application of the formulas uses logarithmic rules, where, for example, multiplication transforms into the addition of logarithm factors: $\log(a\cdot b)=\log(a)+\log(b)$, and then the result of multiplication is $(a\cdot b)=10^{[\log(a)+\log(b)]}$. Today, we do not use logarithmic tables for this type of calculation.

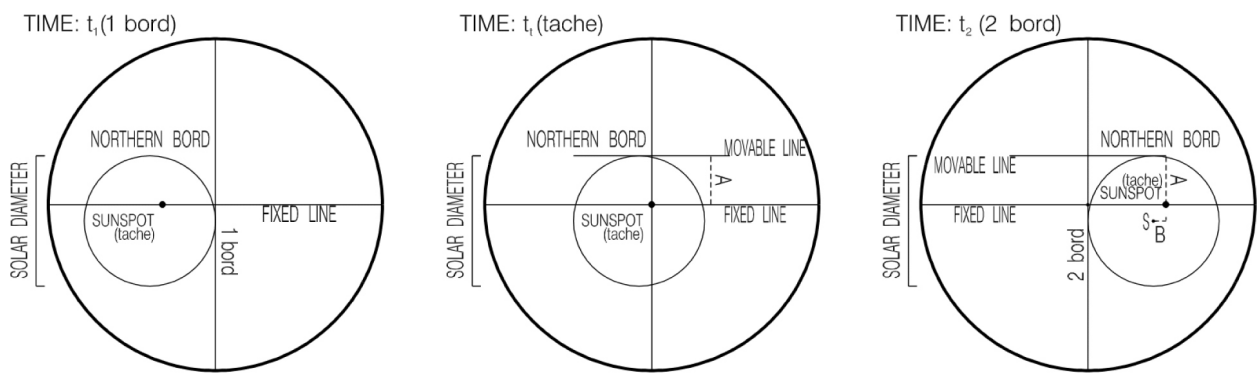

**Figure 1:** The view through the telescope of the solar disk with observed and measured values: A – distance from the northern edge of the solar disk (*bord boreal*); the observed times of passing the vertical line: the time $t_1$ of the 1st edge (*1 bord*); the time $t_t$ of the sunspot (*tache*), the time $t_2$ of the 2nd edge (*2 bord*); and the time difference $B=t_t-(t_1+t_2)/2$ (*Différence*), reconstruction of the description is in *Opuscule II* (Boscovich, 1785, §.I., №20-№26, pages 83-89).

### 3.1. Observations of Ruđer Bošković

Bošković performed the observations in chateau Noslon near Sens, 120 km south of Paris, France in September 1777. The place was equipped by Joseph Jérôme Lefrançois de Lalande (July 11th, 1732, Bourg-en-Bresse, France – April 4th, 1807, Paris, France) with astronomical instruments: telescope, quadrant, etc. (**URL 2**). Bošković was a guest of the cardinal Paul d'Albert de Luynes (January 5th, 1703, Versailles, France – January 21st, 1788, Paris, France), the amateur astronomer, an honorary member of the French Academy of science (*Académie des sciences*) (**URL 3**).

In September 1777, the observations were made by Bošković himself in Sens. He described the methods, formulas, and the example of determination of solar rotation elements: the solar equator inclination i, the longitude of the node Ω, and the period of solar rotation T (**Boscovich, 1785, §.I.-§.XIV., pages 75-169**). In *Opuscule II*, he numbered the paragraphs continuously with Arabic numbers №1 to №165, and chapters with Roman numbers §.I. to §.XIV. *Opuscule II* has the appendix *Appendice* where paragraphs were numbered with Arabic numbers №1 to №20, pages 170-178. There, all the observations he made in September 1777 are given for four sunspots (**Boscovich, 1785, *Appendice*, pages 170-178**).

In *Opuscule II* (**Boscovich, 1785, №4**), his observation procedure was described. Bošković made observations himself using a telescope equipped with a micrometre for the determination of the differences of declination and rectascension of an observed sunspot and a precise pendulum for measuring time (**Préface, №4**).

Bošković's description of the observed sunspot one: *The sunspot was clearly recognizable, regular, of medium size, so that the thread of the telescope could pass through its center*[1]. We can conclude that the sunspot could be type A, B, C, D or G according to *The Zurich Classification System of Sunspot Groups* (**URL 4**).

For six days he observed one medium-sized sunspot in early afternoon about 3 p.m. He observed the sunspot in a series of five measurements. He observed passing times of the solar disk edges of the sunspot, and the vertical distance of the sunspot from the northern edge of the solar disk. He determined the horizontal distance of the sunspot from the solar disk center using passing times of the edges and the sunspot through the vertical line.

In the field of view of the fixed telescope (see **Figure 1**), he observed the passing times of the solar disk $t_1$, the sunspot $t_t$, and the solar disk $t_2$ through the vertical line. At the same time, he measured the vertical position of the sunspot from the northern edge of the solar disk *A*: he put the fixed horizontal line of a micrometre at the sunspot, and then he measured the position of the northern edge of the solar disk with the mobile line of the micrometre. He converted the vertical distance measured using the micrometre into angular seconds using the constant of the micrometre *C*. He determined constant *C* relatively to the apparent diameter of the solar disk described in section *3.2 Astronomic almanac data*. He measured the vertical distance of the sunspot from the solar disk edge in angular seconds using the micrometre.

---

[1] In French original: *La tache étoit bien distincte, régulière, & d'une grandeur médiocre, pour pouvoir faire passer le fil par son milieu.* (**Boscovich, 1785, §. I., №25, page 87**).

**Table 1:** The records of the observed sunspot of six day observations: the 1st line: September 12th, 1777; the 2nd line: north edge (***bord boreal***) with its arithmetic mean the far right (***milieu***); the 3rd line through the 5th lines: the observed times of passing the vertical line: the 1st edge (***1 bord***) - $t_1$, the sunspot (***tache***) – $t_t$, and the 2nd edge (***2 bord***) - $t_2$; and the 6th line: the difference (***Différence***) with its arithmetic mean the far right (***milieu***); September 12th and 15th, 1777 have the wrong values in boxes (**Boscovich, 1785, pages 87-89**).

| | | | | | | | | | | | | | | | | | | Comment: |
|---|---|---|---|---|---|---|---|---|---|---|---|---|---|---|---|---|---|---|
| **12. Sept.** | | | **1777.** | | | | | | | | | | | | | | | |
| bord boreal | | | | 561 | | | 555 | | | 559 | | | 563 | | | 559 | Milieu **559.4** | A |
| | h | ' | " | h | ' | " | h | ' | " | h | ' | " | h | ' | " | | | |
| 1 bord.. | **2** | **59** | **9** | 3 | 6 | 42 | 3 | 10 | 32 | 3 | 14 | 27 | 3 | 22 | 3 | | | $t_1$ |
| tache.. | 3 | 0 | 55 | 3 | 8 | 29 | 3 | 12 | 20 | 3 | 16 | 14 | 3 | 23 | 51 | | | $t_t$ |
| 2 bord.. | 3 | 1 | 16 | 3 | 8 | 50 | 3 | 12 | 40 | 3 | 16 | 35 | **3** | **25** | **12** | | | $t_2$ |
| Différence | | | 42.5 | | | 43.0 | | | 44.0 | | | 43.0 | | | 43.5 | milieu **43.2** | B |
| **13. Sept.** | | | **1777.** | | | | | | | | | | | | | | | |
| bord boreal | | | | 526 | | | 524 | | | 521 | | | 527 | | | 524 | Milieu **524.4** | |
| | h | ' | " | h | ' | " | h | ' | " | h | ' | " | h | ' | " | | | |
| 1 bord.. | **2** | **33** | **4** | 2 | 35 | 44 | 2 | 39 | 48 | 2 | 42 | 33 | 2 | 50 | 11 | | | |
| tache.. | 2 | 34 | 41 | 2 | 37 | 21 | 2 | 41 | 25 | 2 | 44 | 11 | 2 | 51 | 49 | | | |
| 2 bord.. | 2 | 35 | 11 | 2 | 37 | 52 | 2 | 41 | 56 | 2 | 44 | 41 | **2** | **52** | **21** | | | |
| Différence | | | 33.5 | | | 33.0 | | | 33.0 | | | 34.0 | | | 33.0 | milieu **33.3** | |
| **15. Sept.** | | | **1777.** | | | | | | | | | | | | | | | |
| bord boreal | | | | 440 | | | 440 | | | 440 | | | 440 | | | 440 | Milieu **440.0** | |
| | h | ' | " | h | ' | " | h | ' | " | h | ' | " | h | ' | " | | | |
| 1 bord.. | **3** | **6** | **42** | 3 | 14 | 8 | 3 | 17 | 45 | 3 | 20 | 48 | 3 | **34** | 16 | | | |
| tache.. | 3 | 7 | 57 | 3 | 15 | 23 | 3 | 19 | 0 | 3 | 22 | 4 | 3 | 25 | 32 | | | |
| 2 bord.. | 3 | 8 | 50 | 3 | 16 | 15 | 3 | 19 | 53 | 3 | 22 | 56 | **3** | **26** | **24** | | | |
| Différence | | | 11.0 | | | 11.0 | | | 11.0 | | | 12.0 | | | 12.0 | milieu **11.4** | |
| **16. Sept.** | | | **1777.** | | | | | | | | | | | | | | | |
| bord boreal | | | | 388 | | | 388 | | | 389 | | | 390 | | | 388 | Milieu **388.6** | |
| | h | ' | " | h | ' | " | h | ' | " | h | ' | " | h | ' | " | | | |
| 1 bord.. | **3** | **42** | **35** | 3 | 45 | 29 | 3 | 48 | 22 | 3 | 51 | 19 | 3 | 54 | 3 | | | |
| tache.. | 3 | 43 | 39 | 3 | 46 | 33 | 3 | 49 | 25 | 3 | 52 | 23 | 3 | 55 | 8 | | | |
| 2 bord.. | 3 | 44 | 43 | 3 | 47 | 37 | 3 | 50 | 30 | 3 | 53 | 27 | **3** | **56** | **12** | | | |
| Différence | | | 0.0 | | | 0.0 | | | -1.0 | | | 0.0 | | | 1.0 | milieu **0** | |
| **17. Sept.** | | | **1777.** | | | | | | | | | | | | | | | |
| bord boreal | | | | 331 | | | 332 | | | 334 | | | 334 | | | 331 | Milieu **332.4** | |
| | h | ' | " | h | ' | " | h | ' | " | h | ' | " | h | ' | " | | | |
| 1 bord.. | **3** | **18** | **0** | 3 | 24 | 23 | 3 | 28 | 12 | 3 | 31 | 21 | 3 | 35 | 12 | | | |
| tache.. | 3 | 18 | 53 | 3 | 25 | 16 | 3 | 29 | 6 | 3 | 32 | 15 | 3 | 36 | 7 | | | |
| 2 bord.. | 3 | 20 | 7 | 3 | 26 | 30 | 3 | 30 | 19 | 3 | 33 | 29 | **3** | **37** | **20** | | | |
| Différence | | | -10.5 | | | -10.5 | | | -9.5 | | | -10.0 | | | -9.0 | milieu **-9.9** | |
| **19. Sept.** | | | **1777.** | | | | | | | | | | | | | | | |
| bord boreal | | | | 239 | | | 241 | | | 240 | | | 240 | | | 240 | Milieu **240.0** | |
| | h | ' | " | h | ' | " | h | ' | " | h | ' | " | h | ' | " | | | |
| 1 bord.. | **2** | **34** | **26** | 2 | 37 | 14 | 2 | 41 | 51 | 2 | 43 | 47 | 2 | 46 | 43 | | | |
| tache.. | 2 | 35 | 3 | 2 | 37 | 50 | 2 | 42 | 26 | 2 | 44 | 23 | 2 | 47 | 19 | | | |
| 2 bord.. | 2 | 36 | 34 | 2 | 39 | 22 | 2 | 43 | 59 | 2 | 45 | 55 | **2** | **48** | **51** | | | |
| Différence | | | -27.0 | | | -28.0 | | | -29.0 | | | -28.0 | | | -28.0 | milieu **-28** | |

The values: *A* is the distance of the sunspot from the northern edge of the solar disk (*bord boreal*); observed the times of passage of the vertical line (thread, reticule): the time $t_1$ of the 1st edge (*1 bord*), the western (right) edge of the solar disk; the time $t_t$ of the sunspot (*tache*), the time $t_2$ of the 2nd edge (*2 bord*), the eastern (left) edge of the solar disk; and *B* is the time difference $B=t_t-(t_1+t_2)/2$ (see **Figure 1**). The difference *A* is measured with a micrometre, and *B* he determined as the time difference of the sunspot moment $t_t$ and the solar disk centre (*S*) moment $t_S=(t_1+t_2)/2$, $B=t_t-(t_1+t_2)/2=t_t-t_S$. In this paper, the observations are presented in **Table 1 (Bos-**

**Table 2:** Derived observation data: the constant of the micrometre is *C*; and for each date: the observation beginning time: the 1st edge in the 1st series (*1 bord*); the observation ending time: the 2nd edge in the last, the 5st series (*2 bord*); the northern border arithmetic mean *A* (*bord boreal, milieu*); and the difference arithmetic mean *B* (*Différence, milieu*) (**Boscovich, 1785, pages 87-89**).

**Observation data for the calculations**

| C=1915"/1237 units | | | beginning time 1st series (1 bord) | | | ending time 5th series (2 bord) | | | A (bord boreal, milieu) | B (Différence, milieu) |
|---|---|---|---|---|---|---|---|---|---|---|
| Date | | | | | | | | | | |
| day | Month | year | H | M | S | H | M | S | units | " |
| 12 | September | 1777 | 2 | 59 | 9 | 3 | 25 | 12 | 559.4 | 43.2 |
| 13 | September | 1777 | 2 | 33 | 4 | 2 | 52 | 21 | 524.4 | 33.3 |
| 15 | September | 1777 | 3 | 6 | 42 | 3 | 26 | 24 | 440.0 | 11.4 |
| 16 | September | 1777 | 3 | 42 | 35 | 3 | 56 | 12 | 388.6 | 0.0 |
| 17 | September | 1777 | 3 | 18 | 0 | 3 | 37 | 20 | 332.4 | -9.9 |
| 19 | September | 1777 | 2 | 34 | 26 | 2 | 48 | 51 | 240.0 | -28.0 |

**covich, 1785, pages 87-89)**. Bold numbers are the input data derived in **Table 2**. In September 1777, Bošković observed four sunspots (**Boscovich, 1785, *Appendice*, pages 170-178**), but one medium-sized sunspot had the best properties for observation. For determination of the solar rotation elements, he chose sunspot 1 of four which he observed in September 1777. Every day in clear weather, he observed this sunspot in five series and thus obtained homogeneous observations for all six days.

Derived observation data (see **Table 2**) are: the northern border arithmetic mean A (*bord boreal, milieu*); the 1st edge in the 1st series, the observation beginning time (*1 bord*); the 2nd edge in the last, 5th series, the observation ending time (*2 bord*); and the difference arithmetic mean B (*Différence, milieu*), and the constant of the micrometre C. The constant of the micrometre C is determined empirically by observations and data in astronomic almanac (**URL 1, page 108**) described in the section 3.2.

### 3.2. Astronomic almanac data

The solar rotation element reproduction includes: the solar equator inclination *i*, the longitude of the node Ω, and the period of the solar rotation. The astronomic almanac *Connoissance des temps* (**URL 1**) contains all the needed additional data besides his observations for the reproduction of the results in the present work. The astronomic almanac data that Bošković used are: 1. the positions of the Sun and the correction for the mean solar time (see **Table 3**); 2. the longitude of the Sens from Paris (see **Table 4**); 3. the apparent diameter of the Sun (see **Table 5** and **Table 6**); and 4. the inclination of the ecliptic (see **Table 7**).

Daily input data for *Tab. I.* are **boldface** in **Table 3**, later derived in **Table 20**: the solar longitude (*Longitude du Soleil*); the solar declination (*Déclinaison du Soleil*); and Correction for the mean solar time (*Temps moyen au Midi vrai*) (**Connoissance des temps, URL 1, pages 102-103**).

Bošković took the longitude of Sens from Paris (0*h 3m 48s or.) from the Table of meridian differences in hours and degrees from *l'Observatoire Royal de Paris* (see **Table 4**). In the table, the latitudes and the meridian differences have one of two prefixes: * (determined by *Academia*) or † (determined by other astronomers), and suffix *or.* (east of the Paris meridian) or *oc.* (west of the Paris meridian).

Bošković determined the constant of the micrometre C empirically. On September 11th, 1777 with the fixed line of the micrometre, he observed the solar disk edge and with the mobile line of the micrometre, he measured the vertical size of the solar disk. The apparent diameter of the solar disk was measured by a large number of observations. These measurements differed very little from

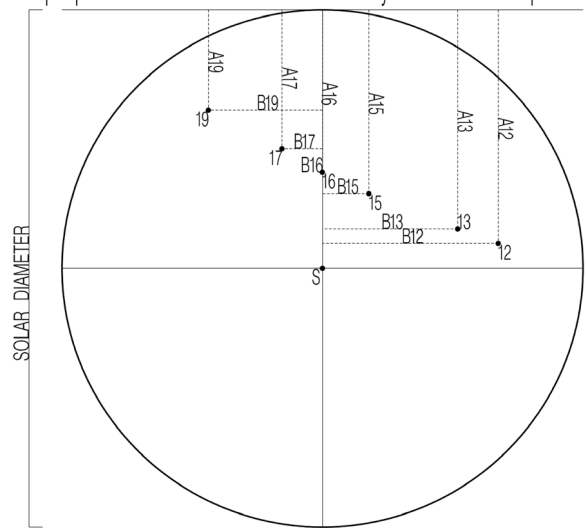

**Figure 2:** Sunspot positions on the solar disk during the period September 12th to 19th, 1777, the view through telescope for all sunspot positions in six days of observation: the sunspot position on the solar disk is the circle with the date and position determined with: A – the distance from the northern solar disk edge, B – the distance from the middle vertical line of the solar disk.

**Table 3:** Daily input data for calculation (**boldface**): solar longitude (*Longitude du Soleil*); solar declination (*Déclinaison du Soleil*); Correction for mean solar time (*Temps moyen au Midi vrai*) (**Connoissance des temps, URL 1, pages 102-103**).

| 102 ÉCLIPSES DE L'ANNÉE 1777. | | | | | | | | | | | | | | | | | | 103 | | | | | | | |
|---|---|---|---|---|---|---|---|---|---|---|---|---|---|---|---|---|---|---|---|---|---|---|---|---|---|
| Jours. | SEPTEMBRE. | COMMENC. du Crép. | | Lever du Soleil. | | Coucher du Soleil. | | FIN du Crépuscule. | | LONGITUDE DU SOLEIL. | | | | Jours. | ASCENSION droite du SOLEIL. | | | DÉCLINAISON du SOLEIL. Boréale. | | | DISTANCE de l'Équinoxe au Soleil. | | | TEMPS MOYEN au Midi vrai. | | | |
| | | H. | M. | H. | M. | H. | M. | H. | M. | S. | D. | M. | S. | | D. | M. | S. | D. | M. | S. | H. | M. | S. | H. | M. | S.D. | Differ. |
| 1 | Lun. S. Leu S. G. | 3 | 20 | 5 | 18 | 6 | 41 | 8 | 39 | 5 | 9 | 16 | 42 | 1 | 160 | 51 | 48 | 8 | 5 | 57 | 13 | 16 | 32 | 11 | 59 | 37.0 | 18.9 |
| 2 | Mardi S. Lazare. | 3 | 22 | 5 | 20 | 6 | 39 | 8 | 37 | 5 | 10 | 14 | 54 | 2 | 161 | 46 | 12 | 7 | 44 | 2 | 13 | 12 | 54 | 11 | 59 | 18.1 | 19.3 |
| 3 | Merc. S. Gregoire | 2 | 25 | 5 | 22 | 6 | 37 | 8 | 34 | 5 | 11 | 13 | 9 | 3 | 162 | 40 | 33 | 7 | 21 | 58 | 13 | 9 | 17 | 11 | 58 | 58.8 | 19.4 |
| 4 | Jeudi. S.ᵉ Marcele | 3 | 27 | 5 | 23 | 6 | 36 | 8 | 32 | 5 | 12 | 11 | 25 | 4 | 163 | 34 | 52 | 6 | 59 | 46 | 13 | 5 | 40 | 11 | 58 | 39.4 | 19.7 |
| 5 | Vend. S. Victorin | 3 | 29 | 5 | 25 | 6 | 34 | 8 | 30 | 5 | 13 | 9 | 43 | 5 | 164 | 29 | 8 | 6 | 37 | 27 | 13 | 2 | 3 | 11 | 58 | 19.7 | 19.8 |
| 6 | Same. S. Onesipe | 3 | 32 | 5 | 27 | 6 | 32 | 8 | 27 | 5 | 14 | 8 | 2 | 6 | 165 | 23 | 16 | 6 | 15 | 2 | 12 | 58 | 26 | 11 | 57 | 59.9 | 20.1 |
| 7 | *Dim.* S. Clou P. | 3 | 34 | 5 | 29 | 6 | 30 | 8 | 25 | 5 | 15 | 6 | 23 | 7 | 166 | 17 | 24 | 5 | 52 | 31 | 12 | 54 | 50 | 11 | 57 | 39.8 | 20.3 |
| 8 | Lundi *Nat. N. D.* | 3 | 36 | 5 | 30 | 6 | 29 | 8 | 23 | 5 | 16 | 4 | 45 | 8 | 167 | 11 | 31 | 5 | 29 | 54 | 12 | 51 | 14 | 11 | 57 | 19.5 | 20.4 |
| 9 | Mardi S. Omer. | 3 | 38 | 5 | 32 | 6 | 27 | 8 | 21 | 5 | 17 | 3 | 9 | 9 | 168 | 5 | 33 | 5 | 7 | 11 | 12 | 47 | 38 | 11 | 56 | 59.1 | 20.5 |
| 10 | Mer. S. Nic. de T. | 3 | 41 | 5 | 34 | 6 | 25 | 8 | 18 | 5 | 18 | 1 | 34 | 10 | 168 | 59 | 31 | 4 | 44 | 24 | 12 | 44 | 2 | 11 | 56 | 38.6 | 20.7 |
| 11 | Jeudi S. Patient. | 3 | 43 | 5 | 36 | 6 | 23 | 8 | 16 | **5** | **19** | **0** | **01** | 11 | 169 | 53 | 27 | **4** | **21** | **31** | 12 | 40 | 26 | 11 | 56 | 17.9 | 20.8 |
| 12 | Ven. S. Serdot. | 3 | 45 | 5 | 37 | 6 | 22 | 8 | 14 | **5** | **19** | **58** | **29** | 12 | 170 | 47 | 23 | **3** | **58** | **33** | 12 | 36 | 50 | **11** | **55** | **57.1** | 20.8 |
| 13 | Sam. S. Maurille. | 3 | 47 | 5 | 39 | 6 | 20 | 8 | 12 | **5** | **20** | **56** | **59** | 13 | 171 | 41 | 19 | **3** | **35** | **32** | 12 | 33 | 15 | **11** | **55** | **36.3** | 20.9 |
| 14 | *Dim.* Exalt. Sᶜ † | 3 | 49 | 5 | 41 | 6 | 18 | 8 | 10 | **5** | **21** | **55** | **31** | 14 | 172 | 35 | 10 | **3** | **12** | **27** | 12 | 29 | 39 | 11 | 55 | 15.4 | 21.0 |
| 15 | Lun. S. Nicodème | 3 | 51 | 5 | 43 | 6 | 16 | 8 | 8 | **5** | **22** | **54** | **04** | 15 | 173 | 29 | 1 | **2** | **49** | **18** | 12 | 26 | 3 | **11** | **54** | **54.4** | 21.0 |
| 16 | Mar. S. Cyprien. | 3 | 54 | 5 | 45 | 6 | 14 | 8 | 5 | **5** | **23** | **52** | **40** | 16 | 174 | 22 | 52 | **2** | **26** | **05** | 12 | 22 | 28 | **11** | **54** | **33.4** | 21.0 |
| 17 | Mercr. *4 Temps*. | 3 | 56 | 5 | 46 | 6 | 13 | 8 | 3 | **5** | **24** | **51** | **17** | 17 | 175 | 16 | 43 | **2** | **02** | **49** | 12 | 18 | 52 | **11** | **54** | **12.4** | 21.0 |
| 18 | Juedi S. Jean Chr. | 3 | 58 | 5 | 48 | 6 | 11 | 8 | 1 | 5 | 25 | 49 | 56 | 18 | 176 | 10 | 35 | 1 | 39 | 31 | 12 | 15 | 17 | 11 | 53 | 51.4 | 21.0 |
| 19 | Vend. S. Janvier. | 4 | 0 | 5 | 50 | 6 | 9 | 7 | 59 | **5** | **26** | **48** | **37** | 19 | 177 | 4 | 28 | **1** | **16** | **12** | 12 | 11 | 41 | **11** | **53** | **30.4** | 20.9 |
| 20 | Same. *vigile-jeûne*. | 4 | 2 | 5 | 52 | 6 | 7 | 7 | 57 | 5 | 27 | 47 | 20 | 20 | 177 | 58 | 21 | 0 | 52 | 50 | 12 | 8 | 6 | 11 | 53 | 9.5 | 20.8 |
| 21 | *Dim. S. Matthieu* | 4 | 4 | 5 | 54 | 6 | 6 | 7 | 55 | 5 | 28 | 46 | 06 | 21 | 178 | 52 | 13 | 0 | 29 | 26 | 12 | 4 | 30 | 11 | 52 | 48.7 | 20.7 |
| 22 | Lun. S. Maurice. | 4 | 6 | 5 | 55 | 6 | 4 | 7 | 53 | 5 | 29 | 44 | 54 | 22 | 179 | 46 | 13 | 0 | 06 | 01 | 12 | 0 | 54 | 11 | 52 | 28.0 | 20.6 |
| | | | | | | | | | | | | | | | | | | Australe | | | | | | | | | |
| 23 | Mardi. S.ᵉ Tecle. | 4 | 8 | 5 | 57 | 6 | 2 | 7 | 51 | 6 | 0 | 43 | 44 | 23 | 180 | 40 | 11 | 0 | 17 | 25 | 11 | 57 | 18 | 11 | 52 | 7.4 | 20.5 |
| 24 | Merc. S. Andoche | 4 | 10 | 5 | 59 | 6 | 0 | 7 | 49 | 6 | 1 | 42 | 36 | 24 | 181 | 34 | 11 | 0 | 40 | 52 | 11 | 53 | 42 | 11 | 51 | 46.9 | 20.2 |
| 25 | Jeudi. S. Firmin. | 4 | 12 | 6 | 1 | 5 | 58 | 7 | 47 | 6 | 2 | 41 | 31 | 25 | 182 | 28 | 13 | 1 | 04 | 18 | 11 | 50 | 6 | 11 | 51 | 26.7 | 20.1 |
| 26 | Vend. S.ᵉ Justine. | 4 | 14 | 6 | 3 | 5 | 57 | 7 | 45 | 6 | 3 | 40 | 28 | 26 | 183 | 22 | 19 | 1 | 27 | 44 | 11 | 46 | 30 | 11 | 51 | 6.6 | 20.0 |
| 27 | Same. S. C. S. D. | 4 | 16 | 6 | 4 | 5 | 55 | 7 | 43 | 6 | 4 | 39 | 28 | 27 | 184 | 16 | 29 | 1 | 51 | 10 | 11 | 42 | 53 | 11 | 50 | 46.6 | 19.7 |
| 28 | *Dim.* S. Ceran. | 4 | 18 | 6 | 6 | 5 | 53 | 7 | 41 | 6 | 5 | 38 | 30 | 28 | 185 | 10 | 42 | 2 | 14 | 35 | 11 | 39 | 16 | 11 | 50 | 26.9 | 19.3 |
| 29 | Lundi *S. Michel*. | 4 | 20 | 6 | 8 | 5 | 51 | 7 | 39 | 6 | 6 | 37 | 34 | 29 | 186 | 4 | 59 | 2 | 37 | 59 | 11 | 35 | 39 | 11 | 50 | 7.6 | 19.2 |
| 30 | Mardi S. Jérôme. | 4 | 22 | 6 | 10 | 5 | 49 | 7 | 37 | 6 | 7 | 36 | 41 | 30 | 186 | 59 | 21 | 3 | 01 | 21 | 11 | 32 | 2 | 11 | 49 | 48.4 | |

Jours décròissent du 1 au 30 de 51'30'' le mat. & de 51'25'' le foir.

each other. He determined the apparent diameter of the solar disk to be *1237* units of the micrometre (**Boscovich, 1785, №22, page 86**).

For the same day, the date September 11th, 1777, we discovered that he made the linear interpolation of the apparent diameter of the Sun. Diameters of the Sun from the **Table 5** are: for September 7th, 1777 diameter is $D_{\odot 7}=31'52.2''$, and for September 13th, 1777 diameter is $D_{\odot 13}=31'55.3''$. The linear interpolation reproduces the apparent solar diameter $D_{\odot}=31'54.26''$ that he rounded in whole seconds $D_{\odot}=31'55''=31\cdot 60''+55''=1915''$ (**URL 1, page 108**). Finally, Bošković determined the constant of the micrometre $C=1915''/1237$ units and $\log C=0.189799$ (see **Table 8**, the second column, the first row).

Another diameter of the Sun in the astronomic almanac is $D_{\odot}=31'57.5''$ (**URL 1, page 260, *DIMENSIONS***

**Table 4:** Table of meridian differences in hours (**boldface**, column en. Temps.) and degrees: Longitude of Sens from Paris: *0ʰ 3ᵐ 48ˢ or.*, the extraction only for Sens (*Connoissance des temps: TABLE DE LA DIFFÉRENCE des Méridiens en heures & degrés, entre l'Observatoire Royal de Paris & les procipaux lieux de la Terre, avec leur latitude ou hauteur de Pole.*) (**URL 1, page 263: The title of the table, and page 268: the longitude of Sens**).

TABLE DE LA
DIFFÉRENCE
DES MÉRIDIENS... 268

| NOMS DES LIEUX. | Différ. des Méridiens | | | | | LATITUDES *ou* Hauteurs du Pole. | | |
|---|---|---|---|---|---|---|---|---|
| | en Temps. | | | en Deg. | | | | |
| | *H.* | *M.* | *S.* | | *D.* | *M.* | *D.* | *M.* | *S.* |
| Sens | 0 | * | 3. | 48. | or. | 0. | 57. | 48 | * | 11. | 56. |

**Table 5:** Apparent diameter of the Sun on September 7th, 13th and 19th, 1777 (**boldface**) in the column *DIAMÉTRE du SOLEIL* (**URL 1, page 108**).

Septembre 1777.                                                                                                                 **108**

| Jours. | TEMPS que le demi-diamèt. du Soleil met à passer par le Mérid. | | DIAMÈTRE du SOLEIL. | | MOUVEM. horaire du SOLEIL. | | LOGARITH. de la distance du SOLEIL. | LIEU du nœud de la LUNE. | | |
|---|---|---|---|---|---|---|---|---|---|---|
|  | Min. | Sec. | Min. | Sec. | Min. | Sec. | La moy. 100000. | S. | D. | M. |
| 1 | 1 | 4.3 | 31 | 49.2 | 2 | 25.4 | 5.003541 | 3 | 15 | 10 |
| **7** | 1 | 4.0 | **31** | **52.2** | 2 | 25.8 | 5.002869 | 3 | 14 | 51 |
| **13** | 1 | 4.0 | **31** | **55.3** | 2 | 26.3 | 5.002149 | 3 | 14 | 32 |
| **19** | 1 | 4.0 | **31** | **58.5** | 2 | 26.8 | 5.001423 | 3 | 14 | 13 |
| 25 | 1 | 4.1 | 32 | 1.7 | 2 | 27.3 | 5.000700 | 3 | 13 | 54 |

**Table 6:** The diameter (**boldface**) of the Sun in the astronomic almanac is 31'57.5" (**URL 1, page 260**).

| *DIMENSIONS DES PLANÈTES, calculées d'après les Observations du PASSAGE DE VÉNUS, qui donnent la parallaxe du Soleil de 8 secondes & demie.* | | | | | |
|---|---|---|---|---|---|
| NOMS des PLANÈTES. | *DIAMÈTRES à la dist. moy. du SOLEIL.* | | *DIAMÈTRES en lieues de 2283 toises* | *DIAMÈTRES par rapport à la TERRE* | *GROSSEUR par rapport à la TERRE* |
| **Le SOLEIL** | *31'* | *57".5* | *323155* | *112.79* | *1435025* |

**Table 7:** The inclination of the ecliptic ε in the astronomic almanac, for the dates: January 1st, April 1st, July 1st, and October 1st, 1777 (**URL 1, page 4, ARTICLES PRINCIPAUX DU CALENDRIER Pour l' Année Commune 1777., OBLIQUITÉ DE L'ÉCLIPTIQUE.**).

| *OBLIIQUITÉ DE L'ÉCLIPTIQUE.* | | | | | | | |
|---|---|---|---|---|---|---|---|
|  | d | ' | " |  | d | ' | " |
| Le 1.er Janv. | 23 | 27 | 46.0 | Le 1.er Juillet | 23 | 27 | 46.9 |
| Le 1.er Avril | 23 | 27 | 46.4 | Le 1.er Oct. | 23 | 27 | 47.3 |

***DES PLANÈTES, calculées d'après les Observations du PASSAGE DE VÉNUS, qui donnent la parallaxe du Soleil de 8 secondes & demie.***).

In the astronomic almanac for the year 1777, there are four values for the inclination of the ecliptic ε, for the dates: January 1st, April 1st, July 1st, and October 1st (**URL 1, page 4, ARTICLES PRINCIPAUX DU CALENDRIER Pour l' Année Commune 1777.**). The inclination of the ecliptic ε=23°28', rounded to whole angular minutes, Bošković used in the *Tab. I.*, the 2nd column, the 14th row (**Boscovich, 1785, §.II., №28, page 90**).

In the next sections, there are Bošković's results in **Table 8** to **Table 19** (**Boscovich, 1785, pages 166-169, Tab. I. to Tab. XII.**), and the present work results are in **Table 22** to **Table 33**.

### 3.3. Bošković's results

Bošković presented his work in 12 tables assigned with Roman numbers. In the tables, the input and the output data are presented in **boldface**. In subsections *3.3. Bošković's results* and *3.4. Present work results,* of the present work, we determined: the inclination *i*, the longitude of the node Ω, and the period of solar rotation. Some tables in the original have no units in the table headers that we added here.

### 3.3.1. Tab. I. and Tab. II.

The first independent part in *Tab. I.* (see **Table 8**) is the determination of the time *T.M.* of the observed sunspot. The second step is the determination of the centre of the solar disk, the solar longitude (*lon.*☉) and the solar declination (*dec.*☉). The centre of the solar disk is the origin for the determination of the sunspot position in the ecliptic coordinate system: the longitude *lon.t* and the latitude *lat.B.t*.

*Tab. I.* (**Boscovich, 1785, page 166**) presents a calculation of the position of the centre of the solar disk, the longitude (lon.☉) and declination (*dec.*☉), and then the position of the sunspot on the solar disk lon.t and lat.B.t[2], and T.M., the last piece of data at the end of the *Tab. I.* The *Tab. I.* is the example for September 12th, 1777 (see **Table 8**). The calculation is repeated for the other 5 days of observation September 13th, 15th, 16th, 17th, and 19th, 1777.

The input data for *Tab. I.* (see **Table 8**) are presented by boldface: 1. the derived observation data in **Table 2** for each day of the observation: the beginning and the

---

[2] For the ecliptic latitude of the sunspot Bošković used two notations: lat.t in the *Tab. I.* and lat.B.t in the *Tab. II.* (**Boscovich, 1785, pages 167 and 168 respectively**).

Table 8: *Tab. I.* (**Boscovich, 1785, page 166**).

| Tab. I. | | | | 12 | Sept. | 1777. | | | | | |
|---|---|---|---|---|---|---|---|---|---|---|---|
| **12ʲ** | **2ʰ** | **59'** | **09"** | C | | | 0.189799 | SI | | | 2.814843 |
| | **3** | **25** | **12** | A= | **559.4** | | 2.747722 | cos.SIB | | | 9.714352 |
| 12 | 3 | 12 | 10 | AxC= | 866.0 | ' ' | 2.937521 | sin.SIB | | | 9.932151 |
| | | -3 | 48 | R= | 15 | 57.4 | | BI | | | 2.529195 |
| 12 | 3 | 8 | 22 | | | 957.4 | | SB | | | 2.746994 |
| | | | | .SB'= | | 91.4 | 8̃.039054 | Fig. 2 | | | |
| | | 58.5 | | D= | 3° | 55'.5 | | .R | | | 7̃.018907 |
| | | 3.1 | | SB' | | 1.5 | | sin.SCI= | 43 | 0 | 9.833750 |
| | | 585 | | cos.D'= | 3 | 57 | 9.998967 | SI= | | 11 | |
| | | 1755 | | | | 15 | 1.176091 | sin.TSC= | 42 | 49 | 9.832288 |
| | | 181.35 | | B= | | 43".2 | 1.635484 | BI | | | 2.529195 |
| | | 30.22 | | B'I | | | 2.810542 | .SI | | | 7̃.185157 |
| | | 7.6 | | tan.B'SI= | 81° | 57' | 0.849596 | sin.CSD= | **20** | **37** | 9.546640 |
| **5ˢ** | **19°** | **58.5** | | cos.I= | 23 | 28 | 9.962508 | .cos.CSD | | | 0̃.028744 |
| 5. | 20. | 6.1 | =lon.☉ | .cos.D= | 3 | 55.5 | 0̃.001020 | SB | | | 2.746994 |
| | | -23.0 | | cos.P'SP= | 23 | 9 | 9.963528 | .SI | | | 7̃.185157 |
| | | 3.1 | | SIB= | 58 | 48 | | sin.TSD= | | 38 24 | 9.793183 |
| | | 230 | | .sin.B'SI | | | 0̃.004301 | lon.†= | 11ˢ | 20 6 | |
| | | 690 | | B'I | | | 2.810542 | lon.t= | 10 | 11 41 | |
| | | 71.30 | | SI= | | 652.9" | 2.814843 | lat.t= | | 20 37 | |
| | | 11.9 | | | | 11' | | T.M. | 12ʲ | 3ʰ .1 | |
| | | 3.0 | | | | | | | | | |
| | **3°** | **58.5** | | | | | | | | | |
| | 3 | 55.5 | =dec.☉ | | | | | | | | |

Inclination of ecliptic: I=23°28'=ε    Apparent solar radius: R=R☉

ending time of daily observations, *A* vertical distance of the sunspot from northern edge of the solar disk determined by the position of the telescope micrometre, *B* the longitude – the time difference of the sunspot from the centre of the solar disk, and the constant of the micrometre *C*; and 2. the astronomic almanac data: The longitude of Sens from Paris 0*ʰ 3ᵐ 48ˢ or. in **Table 4**; the position of the centre of the solar disk, the longitude (lon.☉) and the declination (*dec.*☉), and the correction of time (*Temps moyen au Midi vrai*) in the **Table 3**, the apparent solar diameter *1915"* determined using a linear interpolation of the values in **Table 5**; and the inclination of the ecliptic *ε=23°28'* that Bošković rounded to whole angular minutes from **Table 7**.

The time *T.M.* is the arithmetic mean of the time $t_1$ (*1 bord*) in the first series and the time $t_2$ (*2 bord*) in the fifth (the last) series in a day (see **Table 1**) corrected to the Paris meridian using the time difference of Sens from the astronomic almanac in **Table 4** (**URL 1, pages 263 and 268, the column** *Differ. Des Méridiens, en Temps.*) and then he corrected that true solar time of Paris meridian to the mean solar time using the correction for each day of the observation in **Table 3** (**URL 1, page 103,** *Temps moyen au Midi vrai*). The abbreviation ʲ means franc. jour – day. The time difference of Sens is 3 minutes 48 seconds eastern from Paris. The results of *Tab. I.* (see **Table 8**) for each day of observations are derived in *Tab. II.* (see **Table 9**): the moment of observation *T.M.* and the sunspot positon: the longitude *lon.t* and the latitude *lat.B.t* (**Boscovich, 1785, page 167**).

Table 9: *Tab. II.*: *T.M.* and ecliptic coordinates *lon.t* and *lat.B.t* (**Boscovich, 1785, page 167**).

| Tab. II. | | | | | | | |
|---|---|---|---|---|---|---|---|
| | T.M. | | | lon.t | | | lat.B.t |
| | j | ° | ' | s | ° | ' | ° ' |
| 1 | 12 | 3 | 1 | 10 | 11 | 42 | 20 37 |
| 2 | 13 | 2 | 32 | 10 | 24 | 42 | 20 6 |
| 3 | 15 | 3 | 7 | 11 | 20 | 3 | 19 33 |
| 4 | 16 | 3 | 43 | 0 | 3 | 1 | 19 53 |
| 5 | 17 | 3 | 18 | 0 | 15 | 23 | 21 14 |
| 6 | 19 | 2 | 30 | 1 | 11 | 9 | 22 45 |

### 3.3.2. Tab. III. and Tab. IV.

The longitude of the node Ω Bošković denoted with *N*. From *Tab. II.* (see **Table 9**) he took three positions of

**Table 10:** *Tab. III.* (**Boscovich, 1785, page 167**).

| Tab. III |  | bin. 3 & 5 |  |  |  |
|---|---|---|---|---|---|
|  | s | ° | ' |  | ' |
| .C''-C'= |  | 1 | 41 | = | 101 | ³7.995679 |
| B''-B'= |  | 25 | 20 | = | 1520 | 3.181844 |
| C-C'= |  | 1 | 4 | = | 64 | 1.806180 |
| X= |  | 16 | 3 | = | 963 | 2.983703 |
| **B'=** | 11 | 20 | 3 |  |  |
| L= | 12 | 6 | 6 |  |  |
| **B=** | 10 | 11 | 42 |  |  |
| B+L= | 22 | 17 | 48 |  |  |
| **long.D=** | 11 | 8 | 54 |  |  |
| C-C'= |  |  |  | = | 33 | 1.518514 |
| X= |  | 8 | 17 | = | 497 | 2.696037 |
| **B'=** | 11 | 20 | 3 |  |  |
| L= | 11 | 28 | 20 |  |  |
| **B=** | 10 | 24 | 42 |  |  |
| B+L= | 22 | 23 | 2 |  |  |
| **long.D=** | 11 | 11 | 31 |  |  |

**Table 11:** *Tab. IV.* The longitude of the node *N*=Ω (**Boscovich, 1785, page 167**).

| Tab. IV. | s | ° | ' |  |
|---|---|---|---|---|
|  | 11 | 8 | 54 | 3&5 |
|  |  | 11 | 31 |  |
|  |  | 10 | 43 | 4&5 |
|  |  | 14 | 51 |  |
|  |  | 12 | 14 | 4&6 |
|  |  | 15 | 18 |  |
|  |  | 8 | 18 | 5&6 |
|  |  | 10 | 25 |  |
| somme= |  | 92 | 14 |  |
| **long.D=** | 11 | 11 | 32 |  |
| En ôtant la quatrième & la sixième |  |  |  |  |
| somme= |  | 62 | 5 |  |
| **long.D=** | 11 | 10 | 21 |  |
| **long.N=** | 2 | 10 | 21 |  |

the same sunspot *B, B', B''* (lon.t) and *C, C', C''* (lat.B.t) using his method *2.1. The method for Ω*. The longitude of sunspot culmination D regarding the ecliptic has 3$^s$=90° greater longitude than the ascending node N=Ω=D+3$^s$=D+90°. In *Tab. III.* (see **Table 10**) he determined *D* twice in combination with position 1 before the sunspot culmination regarding the ecliptic and then with position 2 for the pair of sunspot positions 3 and 5 after the culmination, the example for the first pair is given in

---

³ The logarithmic quantity in the denominator is denoted by a point .cos.D, and its value by a tilde above the mantissa (dash in the original example as well as dot . between trigonometric function and argument): .cos.D = 0̃.001020 (**Boscovich, 1785, §.IX., №94, pages 166-169, 125**).

*Tab III* (see **Table 10**) (**Boscovich, 1785, Tab. III. Bin 3 & 5, page 167**).

He made the procedure for another three pairs of the positions of the sunspot and presented them in the *Tab. IV.* (see **Table 11**): the eight D values, the sum, and the arithmetic mean long.D in *Tab. IV.* (see **Table 11**). He discussed the results and decided to remove the 4$^{th}$ and the 6$^{th}$ value, and take into account six other values and he determined another arithmetic mean long.D (**Boscovich, 1785, №114, pages 136-137**). The arithmetic mean long.D increased for *3$^s$* gives the longitude of the node Ω presented in units: sign of Zodiac (*1$^s$=30°*), degrees and minutes N=D+3$^s$=11$^s$10°21'+3$^s$=14$^s$10°21'-12$^s$= 2$^s$10°21' that we converted in angular degrees and minutes *N=(2·30+10)°21'=70°21'*. The final longitude of the node Ω=N=2$^s$10°21'=70°21' Bošković used for determination of the inclination *i*, and the period of the solar rotation.

### 3.3.3. Tab. V. and Tab. VI.

Bošković determined the inclination of the solar equator *i* using the positions of five pairs of one sunspot. The input data are the sunspot positions from *Tab. II.* (see **Table 9**): the longitudes *B, B'* (lon.t), the latitudes BC and B'C' (*lat.B.t*), and the longitude of the node *N* determined in the *Tab. IV.* (see **Table 11**); the output is the inclination of the solar equator *i*. The example for the first pair *B* for the 3$^{rd}$ day and *B'* for the 6$^{th}$ day he presented in the *Tab. V.,* (see **Table 12**). He performed the procedure for the five pairs of the positions presented in the *Tab. VI.* (see **Table 13**) where he presented the sum *38°40'* and the arithmetic mean of the inclination of solar equator *i=7°44'* (**Boscovich, 1785, page 168**).

### 3.3.4. Tab. VII., Tab. VIII., Tab. IX., Tab. X., and Tab. XI.

The determination of the periods of solar rotation uses all the values determined before: *D* in *Tab. IV.* (see **Table 11**), *B* in *Tab. II.* (see **Table 9**), and *i* in *Tab. VI.* (see **Table 13**). *Tab. VII.* (see **Table 14**) and *Tab. VIII.* (see **Table 15**) determine auxiliary values *CP'D* for each day and then in the *Tab. IX.* (see **Table 16**) and in the *Tab. X.* (see **Table 17**) the sidereal period of solar rotation *T'* and then the synodic ones *T''* in *Tab. XI.* (see **Table 18**). The auxiliary value *CP'D* determined in the *Tab. VII.* (see **Table 14**) for six days of one sunspot is in *Tab. VIII.* (see **Table 15**). *Tab. IX.* (see **Table 16**) determines *T'* from six pairs of *T.M.* and the values *CP'D* from *Tab. VIII.* (see **Table 15**). The arithmetic mean of *T'* is the sidereal period of solar rotation. Finally, *Tab. XI.* (see **Table 18**) determines the synodic solar period *T''*.

The calculation of the sidereal and the synodic periods of the solar rotation is performed in two steps: 1. The *CP'D* in *Tab. VII.* (see **Table 14**), and *Tab. VIII.* (see **Table 15**); and 2. The *T'* in the *Tab. IX.* (see **Table 16**) for six pairs of observations using the mean solar time

**Table 12:** *Tab. V.* The inclination of solar equator *i*: input (*N*, *B*, *B'*, *BC*, *B'C'*) and output (*i*) data are **boldface** (**Boscovich, 1785, page 168**).

| Tab. V. | s | ° | ' | | ° | ' | | | ° | ' | | | |
|---|---|---|---|---|---|---|---|---|---|---|---|---|---|
| **N=** | **2** | **10** | **21** | **BC=** | **19** | **33** | | .1.86455 | | | ³9.729426 | cos.BC | 9.974212 |
| **B'=** | **1** | **11** | **9** | **B'C'=** | **22** | **45** | | 0.02015 | | | 8.304275 | sin.DSD' | 9.891115 |
| **B=** | **11** | **20** | **3** | | | | | tan. | 64 | 27 | 0.320529 | sin.D'GF | 9.774899 |
| | | | | SD= | | 0.94235 | | tan. | 1 | 18 | 8.354230 | .C'I | ³1.283329 |
| SD'H= | | 29 | 12 | SD'= | | 0.92220 | | SD'D= | 65 | 45 | | .sin.SD'D | ³0.040119 |
| DSD'= | | 51 | 6 | | | | | SD'H= | 29 | 12 | | **cot.6°12'** | **0.963674** |
| supplém. | | 128 | 54 | | | 1.86455 | | D'GF= | 36 | 33 | | | |
| | | 64 | 27 | | | 0.02015 | | | | | | | |
| | | | | CD= | | 0.33463 | | | | | | | |
| | | | | C'D'= | | 0.38671 | | | | | | | |
| | | | | C'I= | | 0.05208 | | | | | | | |

**Table 13:** *Tab. VI.* The inclination of the solar equator *i* (**Boscovich, 1785, page 168**).

| Tab. VI. | | | | ° | ' |
|---|---|---|---|---|---|
| | 3 | : | 6 | 6 | 12 |
| | 4 | : | 6 | 6 | 22 |
| | 2 | : | 5 | 7 | 28 |
| | 3 | : | 5 | 9 | 26 |
| | 1 | : | 3 | 9 | 12 |
| | | | | 38 | 40 |
| | | | | **7** | **44** |

**Table 14:** *Tab. VII.* (**Boscovich, 1785, page 168**).

| Tab. VII. | s | ° | ' | |
|---|---|---|---|---|
| **long.D=** | **11** | **10** | **21** | |
| **long.B=** | **10** | **11** | **42** | |
| cos.BD= | | 28 | 39 | 9.943279 |
| cot.BC= | | **20** | **37** | 0.424573 |
| tan.PM= | | 66 | 48 | 0.367852 |
| PP'= | | 7 | 44 | |
| .sin.P'M= | | 59 | 4 | ³0.066631 |
| sin.PM | | | | 9.963379 |
| tan.BD | | | | 9.737471 |
| tan.CP'D= | | 30 | 21 | 9.767481 |

**Table 15:** *Tab. VIII.* (Boscovich, 1785, page 168).

| Tab. VIII. | | ° | ' |
|---|---|---|---|
| | 1 | **30** | **21** |
| | 2 | 16 | 36 |
| | 3 | -10 | 17 |
| | 4 | -24 | 0 |
| | 5 | -37 | 6 |
| | 6 | -63 | 56 |

**Table 16:** *Tab. IX.* (Boscovich, 1785, page 168).

| Tab. IX. | | j | | h | |
|---|---|---|---|---|---|
| 4 | | 16 | 3 | 7 | |
| 1 | | 12 | 3 | 0 | |
| | | 4 | 0 | 7 | |
| T | | | | 96.7 | 1.985426 |
| M= | 54° | 21' | = | 3261' | ³6.486649 |
| | | | | | 2.954242 |
| **T'=** | **26ʲ.69** | | | | 1.426317 |

**Table 17:** *Tab. X.* (Boscovich, 1785, page 168).

| Tab. X. | | | Days (ʲ) |
|---|---|---|---|
| **4** | **:** | **1** | **26.69** |
| 5 | : | 1 | 26.75 |
| 6 | : | 1 | 26.65 |
| 5 | : | 2 | 27.04 |
| 6 | : | 2 | 26.82 |
| 6 | : | 3 | 26.67 |
| | | | 160.62 |
| | | | **26.77** |

**Table 18:** *Tab. XI.* (Boscovich, 1785, page 168).

| Tab. XI. | Days (ʲ) | | |
|---|---|---|---|
| A= | 365 | 25 | 2.562590 |
| **T'=** | **26** | **77** | 1.427648 |
| .(A-T')= | 338 | 48 | 470468 |
| **T"=** | **28** | **89** | 1.460706 |

*T.M.* from the second column of *Tab. II.* (see **Table 9**). The arithmetic mean *T'=26.77* days is given in *Tab. X.* (see **Table 17**), and *T"=28.89* days in *Tab. XI.* (see **Ta-** ble 18) using the arithmetic mean of *T'* from *Tab. X.* (see **Table 17**).

### 3.3.5. Tab. XII.

In *Tab. XII.* (see **Table 19**) the calculations of the longitude of the node N and then the inclination of solar equator *i* are presented, using positions of one sunspot in three different sunspot observations. The example for days 1, 3 and 6 is in *Tab. XII.* (see **Table 19**), (**Boscovich, 1785, page 169**). The input data are three positions of the same sunspot longitudes B, B', B'' (*lon.t*), and latitudes BC, B'C', B''C'' (*lat.B.t*) in *Tab. II.* (see **Table 9**). Two calculations with two combinations of three sunspot positions in the upper half of the *Tab. XII.* (see **Table 19**) are equal to the procedure of calculation with two sunspot positions in *Tab. V.* (see **Table 12**). In the furthest right column of the *Tab. XII.* (see **Table 19**), there are four angles *SD'D*, *SD'D''*, *SD''D'*, and *G'D'G* used for further calculation of the longitude of the node N=2$^s$14°03', and the inclination *i*=6°49'.

### 3.4. Present work results

In the present work, we determined the time *T.M.* and the positon of the sunspot for six days of observations in *Tab. I.* (see **Table 22**). The input data are the derived observation data from **Table 2**: the constant of the micrometre *C=1915''/1237*, the inclination of the ecliptic ε=23°28', and the difference from the Paris meridian $\Delta t_{Sens}=3^m 48^s$ *or.*, and the astronomic almanac data for the solar longitude (lon.☉), and the solar declination (*dec.*☉), the correction for the mean solar time (*Temps moyen au Midi vrai*) derived in the **Table 20**.

Determination of the position of the sunspot uses the apparent solar radius $R_☉$ for each day of observation. The

**Table 19:** *Tab. XII.* (**Boscovich, 1785, page 169**).

| Tab. XII. | | | | | | | | | | | |
|---|---|---|---|---|---|---|---|---|---|---|---|
| | | s | ° | ' | | | | | ° | ' | |
| 1 | B= | 10 | 11 | 42 | | SD= | 0.93596 | .1.87831 | | | 9̄.726232 |
| 3 | B'= | 11 | 20 | 3 | | SD'= | 0.94235 | 0.00639 | | | 7.805501 |
| 6 | B''= | 1 | 11 | 9 | | SD''= | 0.92220 | tan. | 70 | 49.5 | 0.458736 |
| | DSD'= | | 38 | 21 | | | 1.87831 | tan. | 0 | 33.7 | 7.990469 |
| | Supplém. | | 141 | 39 | | | 0.00639 | SD'D= | 70 | 16 | |
| | | | 70 | 49.5 | | | 1.86455 | .1.86455 | | | 9̄.729426 |
| | DSD'= | | 51 | 6 | | | 0.02015 | 0.02015 | | | 8.304275 |
| | Supplém. | | 128 | 54 | | CD= | 0.35211 | tan. | 64 | 27 | 0.320529 |
| | | | 64 | 27 | | CD'= | 0.33463 | tan. | 1 | 18 | 8.354230 |
| | BC= | | 20 | 37 | | CD''= | 0.38671 | SD'D''= | 63 | 9 | |
| | B'C'= | | 19 | 33 | | CI= | 0.01748 | SD''D'= | 65 | 45 | |
| | B''C''= | | 22 | 45 | | C''I= | 0.05208 | G'D'G= | 133 | 25 | |
| sin.B'C' | | | | | | 9.524564 | sin.B'C' | | | | 9.524564 |
| cos.BC | | | | | | 9.971256 | cos.B''C'' | | | | 9.964826 |
| sin.DSD' | | | | | | 9.792716 | sin.D''SD' | | | | 9.891115 |
| .CI | | | | | | 1̄.757459 | .C''I' | | | | 1̄.283329 |
| | | | | | | | .sin. | | | | |
| .sin.SD'D | | | | | | 0̄.026284 | SD'D'' | | | | 0̄.049542 |
| D'G= | 11.81 | | | | | 1.072279 | D'G'= | 5.169 | | | 0.713376 |
| D'G'= | 5.169 | | | .16.979 | | 8̄.770088 | cos.B''C'' | | | | 9.964826 |
| | 16.979 | | | 6.641 | ° | ' | 0.822233 | sin.D''SD' | | | 9.891115 |
| | 6.641 | | | tan. | 23 | 17.5 | 9.633969 | sin.D'G'G | | | 9.734353 |
| | | ° | ' | tan. | 9 | 33.5 | 9.226290 | .C''I' | | | 1̄.283329 |
| G'D'G= | | 133 | 25 | D'G'G= | 32 | 51 | | .sin.SD'D | | | 0̄.049542 |
| Suppl. | | 36 | 35 | SD''D'= | 65 | 45 | | | | ° | ' |
| | | 23 | 17.5 | B''SN= | 32 | 54 | | **cot.6°49'** | | 6 | 49 | 0.092217 |
| | | | | | s | ° | ' | | | | |
| | | | | B''= | 1 | 11 | 9 | | | | |
| | | | | N = | 2 | 14 | 3 | | | | |

Bošković put 36, it should be 46, typographic mistake.

**Table 20:** Daily input data for *Tab. I.* calculation: the solar longitude (*Longitude du Soleil*); the solar declination (*Déclinaison du Soleil*); the correction for mean solar time (*Temps moyen au Midi vrai*) (URL 1, pages 102-103).

| Jours. | SEPTEMBRE. | LONGITUDE DU SOLEIL. | | | | DÉCLINAISON du SOLEIL. Boréale. | | | TEMPS MOYEN au Midi vrai. | | | |
|---|---|---|---|---|---|---|---|---|---|---|---|---|
| | | S. | D. | M. | S. | D. | M. | S. | H. | M. | S.D. | Differ. |
| 11 | Jeudi S. Patient. | **5** | **19** | **0** | **01** | **4** | **21** | **31** | **11** | **56** | **17.9** | **20.8** |
| 12 | Ven. S. Serdot | **5** | **19** | **58** | **29** | **3** | **58** | **33** | **11** | **55** | **57.1** | **20.8** |
| 13 | Sam. S. Maurille. | **5** | **20** | **56** | **59** | **3** | **35** | **32** | **11** | **55** | **36.3** | **20.9** |
| 14 | *Dim.* Exalt. S$^c$ † | **5** | **21** | **55** | **31** | **3** | **12** | **27** | **11** | **55** | **15.4** | **21.0** |
| 15 | Lun. S. Nicodème | **5** | **22** | **54** | **04** | **2** | **49** | **18** | **11** | **54** | **54.4** | **21.0** |
| 16 | Mar. S. Cyprien. | **5** | **23** | **52** | **40** | **2** | **26** | **05** | **11** | **54** | **33.4** | **21.0** |
| 17 | Mercr. *4 Temps*. | **5** | **24** | **51** | **17** | **2** | **02** | **49** | **11** | **54** | **12.4** | **21.0** |
| 18 | Juedi S. Jean Chr. | **5** | **25** | **49** | **56** | **1** | **39** | **31** | **11** | **53** | **51.4** | **21.0** |
| 19 | Vend. S. Janvier. | **5** | **26** | **48** | **37** | **1** | **16** | **12** | **11** | **53** | **30.4** | **20.9** |

**Table 21:** Linear interpolation of the apparent solar diameter $D_\odot$ from astronomic almanac given in **boldface**, and then the solar radius $R_\odot = D_\odot/2$ for days of observation in *bold italic* (URL 1, page 108).

| Day | Diameter | Diameter | | Δ" | Radius | |
|---|---|---|---|---|---|---|
| | $D_\odot°$ | ' | " | Δ"/6 | ' | " |
| **7** | **0.531166667** | **31** | **52.2** | **3.1** | **15** | **56.1** |
| 8 | 0.531310185 | 31 | 52.7 | 0.51667 | 15 | 56.4 |
| 9 | 0.531453704 | 31 | 53.2 | 0.51667 | 15 | 56.6 |
| 10 | 0.531597222 | 31 | 53.8 | 0.51667 | 15 | 56.9 |
| 11 | 0.531740741 | 31 | 54.3 | 0.51667 | 15 | 57.1 |
| **12** | 0.531884259 | 31 | 54.8 | 0.51667 | *15* | *57.4* |
| **13** | **0.532027778** | **31** | **55.3** | **3.2** | *15* | *57.7* |
| 14 | 0.532175926 | 31 | 55.8 | 0.53333 | 15 | 57.9 |
| **15** | 0.532324074 | 31 | 56.4 | 0.53333 | *15* | *58.2* |
| **16** | 0.532472222 | 31 | 56.9 | 0.53333 | *15* | *58.5* |
| **17** | 0.532620370 | 31 | 57.4 | 0.53333 | *15* | *58.7* |
| 18 | 0.532768519 | 31 | 58.0 | 0.53333 | 15 | 59.0 |
| **19** | **0.532916667** | **31** | **58.5** | | *15* | *59.3* |

apparent solar diameter for every seventh day in September 1777 is given in the astronomic almanac **Table 5** (**URL 1, page 108**). Linear interpolation of the apparent solar diameter $D_\odot$ determines the daily apparent solar diameter and then divided by two gives us the apparent solar radius $R_\odot$ in **Table 21**. For the days of observation, the apparent radius is given in ***bold italic***.

The observation input data: $t_1$, $t_2$, A, B, and the apparent solar radius $R_\odot$ determined in **Table 21** with position of the Sun (lon.⊙, dec.⊙), and the correction for the mean solar time in **Table 20** are the input data for *Tab. I.* (see **Table 22**).

We reproduced the mean solar time T.M. differently than Bošković did. In the present work, calculation of the solar rotation period uses both results of T.M.: the one Bošković published in *Tab. II.* (see **Table 9**) and the corrected T.M. in *Tab. II.* (see **Table 23**). Both values of T.M. reproduce the solar rotation period twice.

The present work *T.M.*, and the present work sunspot positions *lon.t* and *lat.B.t* in *Tab. II.* (see **Table 23**) are different then the original T.M., and the original sunspot

**Table 22:** *Tab. I.*: the input data: $t_1$, $t_2$, A, B and $R_\odot$ (**Present work**).

Observation data for the calculations

| Date of observation | | | beginning time 1$^{st}$ series (1 bord) | | | ending time 5$^{th}$ series (2 bord) | | | A (bord boreal, milieu) | B (Différence, milieu) | $R_\odot$ Apparent solar radius | |
|---|---|---|---|---|---|---|---|---|---|---|---|---|
| | | | H | M | S | H | M | S | units | " | ' | " |
| 12 | Sept. | 1777. | 2 | 59 | 9 | 3 | 25 | 12 | 559.4 | 43.2 | 15 | 57.4 |
| 13 | Sept. | 1777. | 2 | 33 | 4 | 2 | 52 | 21 | 524.4 | 33.3 | 15 | 57.7 |
| 15 | Sept. | 1777. | 3 | 6 | 42 | 3 | 26 | 24 | 440.0 | 11.4 | 15 | 58.2 |
| 16 | Sept. | 1777. | 3 | 42 | 35 | 3 | 56 | 12 | 388.6 | 0.0 | 15 | 58.5 |
| 17 | Sept. | 1777. | 3 | 18 | 0 | 3 | 37 | 20 | 332.4 | -9.9 | 15 | 58.7 |
| 19 | Sept. | 1777. | 2 | 34 | 26 | 2 | 48 | 51 | 240.0 | -28.0 | 15 | 59.3 |

**Table 23:** *Tab. II.* the present work *T.M.*, and the present work sunspot positions *lon.t* and *lat.B.t* (**Present work**).

| Tab. II. | | | | | | | |
|---|---|---|---|---|---|---|---|
| | T.M. | | | lon.t | | | lat.B.t |
| | j | h | ' | s | ° | ' | ° | ' |
| 1 | 12 | 3 | 4 | 10 | 11 | 42 | 20 | 37 |
| 2 | 13 | 2 | 35 | 10 | 24 | 42 | 20 | 6 |
| 3 | 15 | 3 | 8 | 11 | 20 | 3 | 19 | 33 |
| 4 | 16 | 3 | 40 | 12 | 3 | 6 | 19 | 51 |
| 5 | 17 | 3 | 18 | 12 | 15 | 22 | 21 | 13 |
| 6 | 19 | 2 | 31 | 13 | 11 | 3 | 22 | 44 |

mined *T.M.* with the time correction from true solar time to mean solar time using an astronomical almanac, the furthest right column *Temps moyen au Midi vrai* in the **Table 3** (**URL 1, page 103**).

We determined the periods of the solar rotation *T'*=26.76 days and *T''*=28.87 days (see **Table 35**) using the corrected mean solar time *T.M.* from **Table 34**. Bošković determined periods of the solar rotation *T'*=26.77 days and *T''*=28.89 days using the T.M. values from *Tab. II.* (see **Table 9**). The periods of the solar rotation *T'* and *T''* with *T.M.* determined by Bošković in *Tab. II.* (see **Table 9**), and in the present work in **Table 34**, are almost the same (see **Table 36**).

**Table 24:** *Tab. III.*: The combinations of the sunspot positions and the input data of the original sunspot positions lon.t and lat.B.t and the result *long.D* (**Present work**).

| Observation combinations | | | |
|---|---|---|---|
| 1&3&5 | 1 | 3 | 5 |
| 2&3&5 | 2 | 3 | 5 |
| 1&4&5 | 1 | 4 | 5 |
| 2&4&5 | 2 | 4 | 5 |
| 1&4&6 | 1 | 4 | 6 |
| 2&4&6 | 2 | 4 | 6 |
| 1&5&6 | 1 | 5 | 6 |
| 2&5&6 | 2 | 5 | 6 |
| 1&3&4 | 1 | 3 | 4 |
| 2&3&4 | 2 | 3 | 4 |

| Tab. III. | | lon.t | | lat.B.t | | | lon.t | | lat.B.t | | | long.D | | | | long.D | | |
|---|---|---|---|---|---|---|---|---|---|---|---|---|---|---|---|---|---|---|
| | s | ° | ' | ° | ' | s | ° | ' | ° | ' | | s | ° | ' | | s | ° | ' |
| **1** | 10 | 11 | 42 | 20 | 37 | **2** | 10 | 24 | 42 | 20 | 6 | ◄ Reference sunspot positions 1 & 2 | | | | | | |
| 3 | 11 | 20 | 3 | 19 | 33 | 5 | 0 | 15 | 23 | 21 | 14 | 1&3&5 | 11 | 8 | 54 | 2&3&5 | 11 | 11 | 31 |
| 4 | 0 | 3 | 1 | 19 | 53 | 5 | 0 | 15 | 23 | 21 | 14 | 1&4&5 | 11 | 10 | 43 | 2&4&5 | 11 | 14 | 51 |
| 4 | 0 | 3 | 1 | 19 | 53 | 6 | 1 | 11 | 9 | 22 | 45 | 1&4&6 | 11 | 12 | 14 | 2&4&6 | 11 | 15 | 18 |
| 5 | 0 | 15 | 23 | 21 | 14 | 6 | 1 | 11 | 9 | 22 | 45 | 1&5&6 | 11 | 8 | 18 | 2&5&6 | 11 | 10 | 25 |
| 3 | 11 | 20 | 3 | 19 | 33 | 4 | 0 | 3 | 1 | 19 | 53 | 1&3&4 | 11 | 21 | 37 | 2&3&4 | 11 | 18 | 4 |

positions lon.t and lat.B.t in *Tab. II.* (see **Table 9**). We assume that the results are different because Bošković used the wrong table in an astronomic almanac for the correction of the true solar time to the mean solar time. For the reproduction of the solar rotation elements determination, we used Bošković's original values from *Tab. II.* (see **Table 9**). In that way, we used the same input data as Bošković did and we can compare the results.

### 3.4.1. The mean solar time T.M. and solar rotation periods

Bošković used the values for mean solar time of *T.M.* from *Tab. II.* (see **Table 9**), which are not equal to the values we determined in the present work. We deter-

### 3.4.2. The longitude of the node Ω=N

We determined the longitude of the node Ω (which Bošković assigned as *N*) using his *lon.t* and *lat.B.t* exactly as Bošković did using observations from 1777. The only exception is for the combination *1&3&4*: *long.D(1&3&4)=11$^s$21°37'* (**Present work**), *long.D (1&3&4)=11$^s$21°17'* (**Boscovich 1785, №114, page 137**). We determined standard deviation $\sigma_6$=±1.5058°< $\sigma_8$=±2.5343°<$\sigma_{10}$=±4.2420° and values Δ>>2° for *n*=8 (see **Table 37**).

Modern statistics can eliminate from the results those values that deviate more from the predetermined value. Bošković invented his own L1 fitting method that considers absolute values of differences from arithmetic mean (**Eisenhart, 1961; Ivezić et al., 2014**). He applied normal distribution before Gauß established it in 1809.

Bošković published the *Operas* (**Boscovich, 1785**), before Carl Friedrich Gauß (1777–1855) published the first exposition on the L2 least square fitting method based on the assumption that measurements were distributed by normal distribution as part of the book *Theoria motus corporum coelestium in sectionibus conicis solem ambientium* in 1809 (**Razumović and Triplat Horvat, 2016, 356**).

### 3.5. Bošković's and the present work results

The results we reproduced using the original formulas are very similar to the values that Bošković published in 1785. We can conclude that we successfully reproduced Bošković's example (**Boscovich, 1785, pages 166-169**) in this present work and presented it in **Table 38**.

## 4. Discussion

The first results of the present work reproduced solar rotation elements using input data *T.M.* and sunspot positions *lon.t* and *lat.B.t* that Bošković determined in *Tab. II.* (see **Table 9**) (**Boscovich, 1785, Tab. II., page 167**). The translation of the old-French text revealed a missing link between the observation data and *Tab. I.* (see **Table 8**) that we had in the beginning of the research.

Ruđer Bošković presented a detailed report of all the steps for obtaining the results: the solar equator inclination $i$, the longitude of the node $\Omega$ and period of solar rotation $T$ using logarithmic tables, the astronomical almanac *Connoisance des Temps* (**URL 1**) and his *Opuscule II* (**Boscovich, 1785, №3, pages 76-77**). In September 1777, during a period of six days, in order for the observations of one sunspot to begin, the first step was the determination of the time, and the sunspot position, the longitude and the latitude of the sunspot (**Boscovich, 1785, Tab. II., page 167**). There are many issues resolved during the research: the issues of time, observation input data control, the longitude of the node, apparent diameter of the Sun, the inclination of the ecliptic $\varepsilon$, and positions and mean solar time of the sunspot, missing formulas in some steps.

### 4.1. Time

The time issue should take into account historical epoch, 18th century, when Bošković made the observations and his example. At that time, he used a pendulum and a telescope with a micrometre for precise angle measurements. All the observations were made at about 3 p.m., that means after upper solar culmination, the true solar

**Table 25:** *Tab. IV.* The longitude of the node $N=\Omega$, in the first table using eight ($n=8$) and six ($n=6$) combinations and in the second table using ten ($n=10$) combinations (**Present work**).

| Tab. IV. | s | ° | ' |  |
|---|---|---|---|---|
| 1&3&5 | 11 | 8 | 54 | 338.9014 |
| 2&3&5 | 11 | 11 | 31 | 341.5136 |
| 1&4&5 | 11 | 10 | 43 | 340.7172 |
| **2&4&5** | **11** | **14** | **51** | **344.8507** |
| 1&4&6 | 11 | 12 | 14 | 342.2359 |
| **2&4&6** | **11** | **15** | **18** | **345.2994** |
| 1&5&6 | 11 | 8 | 18 | 338.3034 |
| 2&5&6 | 11 | 10 | 25 | 340.4146 |
| ∑ | 91 | 2 | 14 | 2732.236 |
| (8) long.D= | 11 | 11 | 32 | 341.5295 |
| ∑ | 68 | 02 | 05 | 2042.086 |
| (6) long.D= | 11 | 10 | 21 | 340.3477 |
| **long.N=** | **2** | **10** | **21** | **70.34767** |

| Tab. IV. | s | ° | ' |  | ° |
|---|---|---|---|---|---|
| 1&3&5 | 11 | 08 | 54 |  | 338.9014 |
| 2&3&5 | 11 | 11 | 31 |  | 341.5136 |
| 1&4&5 | 11 | 10 | 43 |  | 340.7172 |
| 2&4&5 | 11 | 14 | 51 |  | 344.8507 |
| 1&4&6 | 11 | 12 | 14 |  | 342.2359 |
| 2&4&6 | 11 | 15 | 18 |  | 345.2994 |
| 1&5&6 | 11 | 08 | 18 |  | 338.3034 |
| 2&5&6 | 11 | 10 | 25 |  | 340.4146 |
| 1&3&4 | 11 | 21 | 37 |  | 351.6217 |
| 2&3&4 | 11 | 18 | 04 |  | 348.0725 |
|  | 114 | 11 | 56 |  | 3431.93 |
|  |  |  |  | n= | 10 |
| **(10) long.D=** | **11** | **13** | **12** |  | **343.193** |

**Table 26:** *Tab. V.* The input data *lon.t* and *lat.B.t*, and the results for the solar equator inclination *i* (Present work).

| Tab. V. |  | lon.t | | | lat.B.t | | | lon.t | | | lat.B.t | | | i | |
|---|---|---|---|---|---|---|---|---|---|---|---|---|---|---|---|
|  |  | s | ° | ' | ° | ' |  | s | ° | ' | ° | ' |  | ° | ' |
|  | 3 | 11 | 20 | 3 | 19 | 33 | 6 | 1 | 11 | 9 | 22 | 45 |  | **06** | **12** |
|  | 4 | 0 | 3 | 1 | 19 | 53 | 6 | 1 | 11 | 9 | 22 | 45 |  | **06** | **22** |
|  | 2 | 10 | 24 | 42 | 20 | 6 | 5 | 0 | 15 | 23 | 21 | 14 |  | **07** | **28** |
|  | 3 | 11 | 20 | 3 | 19 | 33 | 5 | 0 | 15 | 23 | 21 | 14 |  | **09** | **26** |
|  | 1 | 10 | 11 | 42 | 20 | 37 | 3 | 11 | 20 | 3 | 19 | 33 |  | **09** | **14** |
| N= | 2 | 10 | 21 | ◄ The longitude of the node N=Ω | | | | | | | The inclination i ▲ | | | | |

**Table 27:** *Tab. VI.* Sunspot position pairs and the solar equator inclination *i* (**Present work**).

| Tab. VI. | ° | ' | ° |
|---|---|---|---|
| 3:6 | 6 | 12 | 6.205398 |
| 4:6 | 6 | 22 | 6.363927 |
| 2:5 | 7 | 28 | 7.473499 |
| 3:5 | 9 | 26 | 9.437862 |
| 1:3 | 9 | 14 | 9.238984 |
|  | 38 | 43 | 38.71967 |
| i= | 7 | 45 | 7.743934 |

**Table 28:** *Tab. VII.* The input data *long.D* and the sunspot position *lon.t, lat.B.t* and the result the angle *CP'D* (**Present work**).

| Tab. VII. | lon.t | | | lat.B.t | | CP'D | |
|---|---|---|---|---|---|---|---|
|  | s | ° | ' | ° | ' | ° | ' |
| 1 | 10 | 11 | 42 | 20 | 37 | **30** | **21** |
| 2 | 10 | 24 | 42 | 20 | 48 | **16** | **36** |
| 3 | 11 | 20 | 3 | 19 | 33 | **-10** | **17** |
| 4 | 0 | 3 | 1 | 19 | 53 | **-23** | **60** |
| 5 | 0 | 15 | 23 | 21 | 14 | **-37** | **06** |
| 6 | 1 | 11 | 9 | 22 | 45 | **-63** | **56** |
| long.D= | **11** | **10** | **21** | **07** | **45** | =i |  |

**Table 29:** *Tab. VIII.* The angle *CP'D* (**Present work**).

| Tab. VIII | ° | ' | ° |
|---|---|---|---|
| 1 | 30 | 21 | 30.3443 |
| 2 | 16 | 36 | 16.5969 |
| 2 | -10 | 17 | -10.2827 |
| 4 | -23 | 60 | -23.9988 |
| 4 | -37 | 06 | -37.1005 |
| 6 | -63 | 56 | -63.9365 |

**Table 30:** *Tab. IX.* The input data *T.M.* of the sunspot pair, *CP'D* and the results for the solar rotation period *T'* (**Present work**).

| | T.M. | | | | T.M. | | | CP'D | | T' |
|---|---|---|---|---|---|---|---|---|---|---|
| | j | h | m | | j | h | m | ° | ' | (days) |
| **4** | 16 | 3 | 43 | **1** | 12 | 3 | 1 | 30 | 21 | 26.68813 |
| **5** | 17 | 3 | 18 | **1** | 12 | 3 | 1 | 16 | 36 | 26.74944 |
| **6** | 19 | 2 | 30 | **1** | 12 | 3 | 1 | -10 | 17 | 26.64575 |
| **5** | 17 | 3 | 18 | **2** | 13 | 2 | 32 | -23 | 60 | 27.02980 |
| **6** | 19 | 2 | 30 | **2** | 13 | 2 | 32 | -37 | 6 | 26.81498 |
| **6** | 19 | 2 | 30 | **3** | 15 | 3 | 7 | -63 | 56 | 26.66822 |

**Table 31:** *Tab. X.* (**Present work**)

| Tab. X. | days | days |
|---|---|---|
| 4:1 | 26.69 | 26.68813 |
| 5:1 | 26.75 | 26.74944 |
| 6:1 | 26.65 | 26.64575 |
| 5:2 | 27.03 | 27.02980 |
| 6:2 | 26.81 | 26.81498 |
| 6:3 | 26.67 | 26.66822 |
|  | 160.60 | 160.5963 |
|  | 6 | 6 |
|  | 26.77 | 26.76605 |

**Table 32:** *Tab. XI.* (**Present work**)

| Tab. XI. | days | | days |
|---|---|---|---|
| A= | 365 | 25 | 365.25 |
| T'= | 26 | 77 | 26.77 |
| (A-T')= | 338 | 48 | 338.48 |
| T''= |  |  | 28.89 |

noon. From the true solar noon, Bošković could measure all the time moments in his observation tables using the pendulum as he mentioned in his work (**Boscovich, 1785, №4, pages 77-78**). He accomplished 6 day records of the one sunspot with five series of observations with three items of time: western edge of the solar disk (*1 bord*), the sunspot (*tache*), and the eastern edge of the solar disk (*2 bord*), and in the first line units of micrometre (*bord boréal*), and the last line the difference (*Différence*). He determined the centre of solar disk as arithmetic mean of left (western) and right (eastern) solar disk edge.

### 4.2. Observation input data control

The six day observations of the sunspot (**Boscovich, 1785, pages 87-89**) were put into a spreadsheet (see **Table 39**) where we made data input control (see **Table 40**): 1. arithmetic means (*milieu*) for the micrometre data *A* (*bord boréal*), and time differences *B* (*Différences*), 2. times of observed moments $t_1$, $t_t$, and $t_2$ (*1 bord, tache, 2 bord*) should be ascending $t_1<t_t<t_2$, and the time differences should be approximately constant, 3. the time difference $B=t_t-(t_1+t_2)/2$. We made differences in each of five series for each day $\Delta_1$, $\Delta_2$, $\Delta_3$, between the series $\Delta_4$, and the duration of the daily observations $\Delta_5$. For sunspot 1, we consulted all the observations in the *Appendice* (**Boscovich, 1785, pages 170-178**). For sunspot 1, we made the control of data input 1, 2, and 3.

$\Delta_1 = t^i_{tache} - t^i_{1\ bord}$

$\Delta_2 = t^i_{2\ bord} - t^i_{tache}$

$\Delta_3 = t^i_{2\ bord} - t^i_{1\ bord}$

$\Delta_4 = t^{i+1}_{1\ bord} - t^i_{2\ bord}$

$\Delta_5 = t^5_{2\ bord} - t^1_{1\ bord}$

$\Delta_1$ and $\Delta_2$ – the time difference of the neighbouring time observations of the same series

$\Delta_3$ – the time difference of the time observations of one series

**Table 33:** *Tab. XII.* The longitude of the node *N=Ω* and the solar equator inclination *i* (Present work).

**Tab. XII. 1&3&6**

| | | s | ° | ' | | | | | ° | ' | |
|---|---|---|---|---|---|---|---|---|---|---|---|
| 1 | B= | 10 | 11 | 42 | 311.7000 | SD= | | 0.93596 | 1.87831 | | |
| 3 | B'= | 11 | 20 | 3 | 350.0500 | SD'= | | 0.94235 | 0.00639 | | |
| 6 | B''= | 1 | 11 | 9 | 401.1500 | SD''= | | 0.92220 | tan | 70 | 49.5 | 70.8250 |
| | DSD'= | | 38 | 21 | 38.3500 | | | 1.87831 | tan | 00 | 33.6 | 0.5607 |
| | Supplém. | | 141 | 39 | 141.6500 | | | 0.00639 | SD'D= | 70 | 15.9 | 70.2643 |
| | | | 70 | 49.5 | 70.8250 | | | 1.86455 | 1.86455 | | |
| | D''SD'= | | 51 | 06 | 51.1000 | | | 0.02015 | 0.02015 | | |
| | Supplém. | | 128 | 54 | 128.9000 | CD= | | 0.35211 | tan | 64 | 27 | 64.4500 |
| | | | 64 | 27 | 64.4500 | C'D'= | | 0.33463 | tan | 01 | 18 | 1.2950 |
| 1 | BC= | | 20 | 37 | 20.6167 | C''D''= | | 0.38671 | SD'D''= | 63 | 09 | 63.1550 |
| 3 | B'C'= | | 19 | 33 | 19.5500 | CI= | | 0.01748 | SD''D= | 65 | 45 | 65.7450 |
| 6 | B''C''= | | 22 | 45 | 22.7500 | C''I'= | | 0.05208 | G'D'G= | 133 | 25 | 133.4193 |
| | sin.B'C' | | | | 0.33463 | | | | sin.B'C' | | | 0.33463 |
| | cos.BC | | | | 0.93596 | | | | cos.B''C'' | | | 0.92220 |
| | sin.DSD' | | | | 0.62046 | | | | sin.D''SD' | | | 0.77824 |
| | .CI | | | | 0.01748 | | | | .C''I' | | | 0.05208 |
| | .sin.SD'D | | | | 0.94126 | | | | .sin.SD'D'' | | | 0.89223 |
| | D'G= | | | | 11.80787 | | | | D'G'= | | | 5.16824 |
| | D'G'= | | | | 5.16824 | | | 16.97611 | cos.B''C'' | | | 0.92220 |
| | | | | | 16.97611 | | ° | ' | 6.63963 | sin.D''SD' | | | 0.77824 |
| | | | | | 6.63963 | tan. | 23 | 17.4 | 23.29035 | sin.D'G'G | | | 0.54240 |
| | | | | | | tan. | 09 | 33.4 | 9.55686 | .C''I' | | | 0.05208 |
| | G'D'G= | | 133 | 25 | 133.4193 | D'G'G= | 32 | 51 | 32.84721 | .sin.SD'D'' | | | 0.89223 |
| | Suppl. | | 46 | 35 | 46.5807 | SD''D'= | 65 | 45 | 65.74495 | tan.i | | | 8.37746 |
| | | | 23 | 17.4 | 23.2903 | B''SN= | 32 | 54 | 32.89774 | cot.i | ° | ' | 0.11937 |
| Bošković put 36, it should be 46, typographic mistake. | | | | | | | s | ° | ' | cot.6°48' | 06 | 48 | 6.807069 |
| | | | | | | B''= | 1 | 11 | 9 41.15000 | | | |
| | | | | | | Ω = N = | 2 | 14 | 3 74.04774 | **I =** | **06** | **48** | |

Δ$_4$ – the time difference of the time observations of the neighbouring series

Δ$_5$ – the time difference of the beginning time of the first series and the ending time of the last (fifth) series

We present the example for the 1st series on September 12th, 1777:

Δ$_1$ = 3$^h$00'55" – 2$^h$59'09" = 106"

Δ$_2$ = 3$^h$01'16" – 3$^h$00'55" = 21"

Δ$_3$ = 3$^h$01'16" – 2$^h$59'09" = 127"

Δ$_4$ = 3$^h$06'42" – 3$^h$01'16" = 326"

Δ$_5$ = 3$^h$25'12" – 2$^h$59'09" = 1563"

On September 12th, 1777, the third time in the fifth series of 3$^h$25'12" has a difference of Δ$_3$=189". That time should be 3$^h$24'12", since this difference is approximately 60 seconds longer than the time differences recorded in the other four series of that day, *127"* and *128"*. That confirms the time difference that Bošković has in the table of observation *43.5"* of the sunspot moment $t_t$ from the solar disk centre *S*, $t_S=(t_1+t_2)/2$, $B=t_t-(t_1+t_2)/2=t_t-t_S$=3$^h$23'51"–(3$^h$22'03"+3$^h$24'12")/2=43.5". In the fifth series of observation, the time $t_2$=3$^h$25'12" gives the wrong time difference *B=13.5"*. Bošković used the mentioned time item in *Tab. I.* (see **Table 8**). The *Appendice* contains the same value for the September 12th, 1777 in the fifth series of observation $t_2=3^h25'12''$. (**Boscovich, 1785,** *Appendice,* №4, page 171).

**Table 34:** Time *T.M.*, using equation of time (**URL 1, page 103**, *Temps moyen au Midi vrai*).

| T.M. | | | |
|---|---|---|---|
| j | h | ' | '' |
| 12 | 3 | 1 | 2 |
| 13 | 2 | 13 | 2 |
| 15 | 3 | 3 | 0 |
| 16 | 3 | 16 | 0 |
| 17 | 3 | 7 | 1 |
| 19 | 2 | 12 | 1 |

**Table 35:** *Tab. X.* and *Tab. XI.* with *T.M.* using equation of time in the **Table 34** (**Present work**).

| Tab. X. | Days |
|---|---|
| 4:1 | 26.66 |
| 5:1 | 26.74 |
| 6:1 | 26.64 |
| 5:2 | 27.02 |
| 6:2 | 26.81 |
| 6:3 | 26.67 |
|  | 160.54 |
|  | 26.76 |

| Tab. XI. | days | | days |
|---|---|---|---|
| A= | 365 | 25 | 365.25 |
| T'= | 26 | 76 | 26.76 |
| (A-T')= | 338 | 48 | 338.49 |
| T''= | | | **28.87** |

**Table 36:** The sidereal and synodic periods of solar rotation using the original Bošković's *T.M.* and in present work corrected *T.M.*

| Solar rotation period | Sidereal | Synodic |
|---|---|---|
|  | T' (days) | T'' (days) |
| original Bošković's T.M. | 26.77 | 28.89 |
| present work corrected T.M. | 26.76 | 28.87 |

On September 15th, 1777 we found the negative difference $\Delta_3=-472''$, where is a typographical mistake, the value $3^h34'16''$ should be $3^h24'16''$, which confirms the same value in the *Appendice* (**Boscovich, 1785,** *Appendice,* №7, page 172).

On September 19th, 1777 the difference between series 3 and 4 was negative $\Delta_4=-12''$. That cannot be real. The first time $t_1$ is near $2^h42'00''$ and the last time $t_2$ is near $2^h43'00''$, and we could presume that the times in the third series could be one minute less ($t_1=2^h40'51''$, $t_t=2^h41'26''$ and $t_2=2^h42'59''$) or two minutes less ($t_1=2^h39'51''$, $t_t=2^h40'26''$ and $t_2=2^h41'59''$). These presumed values give us respectively $\Delta_4=48''$, and $\Delta_4=108''$, and the same difference $B=-29$. We presume that the one minute less values are more probable because $\Delta_4$ is near to the first $\Delta_4=40''$, and the fourth $\Delta_4=48''$ values. In the *§.I.* and in the *Appendice* for September 19th, 1777, the values are the same (**Boscovich, 1785,** *§.I.,* №26, page 89, *Appendice,* №11, page 173). These values we do not use for the present work example reproduction.

Daily observation duration $\Delta_5$ is from $26'03''<\Delta_5<13'37''$, approximately $\Delta_5=20'±6'$.

### 4.3. The longitude of the node N=Ω

Bošković discussed the differences of the 8 values arithmetic mean of *long.D* in *Tab. IV.* (see **Table 11**). He identified the differences of the 4th and the 6th values that are too far from the others, more than *2°*. The arithmetic mean of six other values $D=11^s10°21'$, and new differences were less than *2°* (see **Table 37**). The final longitude of the node $\Omega=N=2^s10°21'=70°21'=N$. Furthermore, Bošković added *long.D* pair 3&4 with values *21°17'* and *18°04'*. The new total sum is of *131°45'*, divided by 10 and the arithmetic mean is $D=11^s13°09'$. The longitude of the node using 10 values is $N=2^s13°09'=73°09'$. He concluded that the result is very near to the longitude of the node through three points $N=2^s14°03'$ in *Tab. XII.* (see **Table 19**) (**Boscovich, 1785,** №115, page 137).

**Table 37:** The longitude of the sunspot culmination *long.D* (Ω=N=long.D±90°) (Bošković assigned *N*), differences from arithmetic means with six, eight, and 10 values, standard deviation σ, the values *Δ>>2°* for *n=8* are boldfaced.

| Tab. IV. | s | ° | ' | ° | n=6 (°) | Δ (°) | Δ² | n=8 (°) | Δ (°) | Δ² | n=10 (°) | Δ (°) | Δ² |
|---|---|---|---|---|---|---|---|---|---|---|---|---|---|
| 1&3&5 | 11 | 8 | 54 | 338.9014 | 338.9014 | 1.4463 | 2.0917 | 338.9014 | 2.6281 | 6.9070 | 338.9014 | 4.2916 | 18.4182 |
| 2&3&5 | 11 | 11 | 31 | 341.5136 | 341.5136 | -1.1659 | 1.3594 | 341.5136 | 0.0159 | 0.0003 | 341.5136 | 1.6794 | 2.8205 |
| 1&4&5 | 11 | 10 | 43 | 340.7172 | 340.7172 | -0.3695 | 0.1365 | 340.7172 | 0.8123 | 0.6599 | 340.7172 | 2.4758 | 6.1298 |
| 2&4&5 | 11 | 14 | 51 | 344.8507 |  |  |  | 344.8507 | **-3.3212** | 11.0302 | 344.8507 | -1.6577 | 2.7478 |
| 1&4&6 | 11 | 12 | 14 | 342.2359 | 342.2359 | -1.8882 | 3.5654 | 342.2359 | -0.7064 | 0.4990 | 342.2359 | 0.9571 | 0.9161 |
| 2&4&6 | 11 | 15 | 18 | 345.2994 |  |  |  | 345.2994 | **-3.7699** | 14.2120 | 345.2994 | -2.1064 | 4.4368 |
| 1&5&6 | 11 | 8 | 18 | 338.3034 | 338.3034 | 2.0443 | 4.1791 | 338.3034 | 3.2261 | 10.4079 | 338.3034 | 4.8896 | 23.9086 |
| 2&5&6 | 11 | 10 | 25 | 340.4146 | 340.4146 | -0.0669 | 0.0045 | 340.4146 | 1.1149 | 1.2431 | 340.4146 | 2.7784 | 7.7197 |
| 1&3&4 | 11 | 21 | 37 | 351.6217 |  |  |  |  |  | 0.0000 | 351.6217 | -8.4287 | 71.0423 |
| 2&3&4 | 11 | 18 | 4 | 348.0725 |  |  |  |  |  | 0.0000 | 348.0725 | -4.8795 | 23.8091 |
|  | 114 | 11 | 56 | 3431.93 | 2042.0861 | 0.0000 | 11.3366 | 2732.2362 | 0.0000 | 44.9592 | 3431.9304 | 0.0000 | 161.9489 |
|  |  |  |  | n= | 10 | 6 | = n | σ₆= | 8 | = n | σ₈= | 10 | = n | σ₁₀= |
| long.D= | 11 | 13 | 12 | 343.193 | 340.34768 | ± | 1.5058 | 341.52953 | ± | 2.5343 | 343.19304 | ± | 4.2420 |
|  |  |  | Ω=N=long.D±90°= | 70°20'52" | ± | 1°30'21" | 71°31'46" | ± | 2°32'04" | 73°11'35" | ± | 4°14'31" |

**Table 38:** The longitude of the node *N=Ω*, the solar equator inclination *i*, and the period of solar rotation *T'* and *T"* (**Boscovich, 1785, and present work**).

| | Boscovich (1785) | | | | | Present work | | | | | |
|---|---|---|---|---|---|---|---|---|---|---|---|
| | N=Ω | | *i* | | *T'* | *T"* | | N=Ω | | *i* | | *T'* | *T"* | |
| | ° | ' | ° | ' | days | days | | ° | ' | ° | ' | days | days | |
| Number of pairs: | *Tab. IV.* | | *Tab. VI.* | | *Tab. X.* | *Tab. XI.* | Number of pairs: | *Tab. IV.* | | *Tab. VI.* | | *Tab. X.* | *Tab. XI.* | T.M. from: |
| 5 pairs | | | *7* | *44* | | | 5 pairs | | | *7* | *45* | | | |
| 6 pairs | *70* | *21* | | | *26.77* | *28.89* | 6 pairs | *70* | *21* | | | *26.77* | *28.89* | Original T.M. (Boscovich, 1785) |
| | | | | | | | 6 pairs | | | | | *26.76* | *28.87* | Present work T.M. |
| 8 pairs | *71* | *32* | | | | | 8 pairs | *71* | *32* | | | | | |
| 10 pairs | *73* | *09* | | | | | 10 pairs | *73* | *12* | | | | | |
| Sunspot positions: | *Tab. XII.* | | *Tab. XII.* | | | | Sunspot positions: | *Tab. XII.* | | *Tab. XII.* | | | | |
| 1, 3 & 6 | *74* | *03* | *6* | *49* | | | 1, 3 & 6 | *74* | *03* | *6* | *48* | | | |

**Table 39:** The record of the observed sunspot of one day observations: the 1st line September 12th, 1777; the 2nd line: north edge (***bord boreal***) with its arithmetic mean the far right (***milieu***); the 3rd line through the 5th lines, the observed times of passing the vertical line: the 1st edge (***1 bord***), the sunspot (***tache***), the 2nd edge (***2 bord***); and the difference (***Différence***) with its arithmetic mean at the far right (***milieu***), (**Boscovich, 1785, page 87**).

**12. Sept. 1777.**

| bord boreal | | | 561 | | | 555 | | | 559 | | | 563 | | | 559 | milieu | **559.4** | A |
|---|---|---|---|---|---|---|---|---|---|---|---|---|---|---|---|---|---|---|
| | h | ' | " | h | ' | " | h | ' | " | h | ' | " | h | ' | " | | | |
| 1 bord.. | **2** | **59** | **9** | 3 | 6 | 42 | 3 | 10 | 32 | 3 | 14 | 27 | 3 | 22 | 3 | | | $t_1$ |
| tache.. | 3 | 0 | 55 | 3 | 8 | 29 | 3 | 12 | 20 | 3 | 16 | 14 | 3 | 23 | 51 | | | $t_t$ |
| 2 bord.. | 3 | 1 | 16 | 3 | 8 | 50 | 3 | 12 | 40 | 3 | 16 | 35 | **3** | **25** | **12** | | | $t_2$ |
| Différence | | | 42.5 | | | 43.0 | | | 44.0 | | | 43.0 | | | 43.5 | milieu | **43.2** | B |

### 4.4. Apparent diameter of the Sun

For determination of all the sunspot positions, we used the constant of micrometre *C*. Bošković determined the constant *C* empirically. He observed the solar disk multiple times from the northern to the southern edge of the visible diameter, measuring in the units of micrometre. The apparent solar diameter for that day he determined by linear interpolation of the apparent diameters from the astronomic almanac. He determined the constant *C* for that day. For all the days, he used this constant. The apparent solar diameter changes on a daily basis, as we can conclude from **Table 5** (**URL 1, page 108**). The difference of the diameters in the observation period (September 12th to 19th, 1777) is: $\Delta D_{\odot 12}^{19}=D_{\odot 19}-D_{\odot 12}=31'58.5''-31'55.3''=3.2''$. For other days, the measured diameter would be larger proportionally in the units of micrometre, so the constant of the micrometre would be approximately the same.

On September 11th, 1777, Bošković used an optical micrometre to measure the apparent diameter of the Sun in many repetitions of 1237 units. By interpolation for that day, September 11th, 1777, we determined the apparent diameter of the Sun *31'54.2666"* rounded to one decimal *31'54.3"* or to whole seconds *31'54"* and for September 12th, 1777 we get *31'54.7833"* rounded to one decimal *31'54.8"* or to whole seconds *31'55"* (see **Table 5**). Bošković determined the diameter of the Sun for the same day, and he calculated the constant C with the data for September 12th, 1777 $D_\odot=31'55''=31\cdot 60''+55''=1915''$, which is valid for the day after Bošković performed that observation (**Boscovich, 1785, №22, page 86**). Later in his text (**Boscovich, 1785**), he no longer deals with this but reckons with *logC*. By logarithming the expression C=1915/1237, Bošković uses *logC=0.189799*. By arithmetic check, we have *log1915–log1237=0.189799078* which is within the order of magnitude.

Given that the values in the astronomical almanac are to one decimal place, for September 11th, 1777, it would be correct to calculate C=1914.3/1237 or *logC= log1914.3–log1237=0.189640299*. The relative error of Bošković's diameter $D_{\odot B}$ and correctly interpolated $D_{\odot D}$

$$\Delta D_\odot=(D_{\odot B}-D_{\odot D})/D_{\odot D}=$$
$$=(1915-1914.3)/1914.3=0.7/1914.3=$$
$$=0.000365668\approx 0.04\%$$

$\Delta D_\odot=0.04\%$ is not significant.

### 4.5. The inclination of the ecliptic ε

Bošković used the inclination of the ecliptic *ε=23°28'*. He had more precise values in the astronomic almanac for four days in the year 1777 in **Table 7 (URL 1, page 4)**. The inclination could be linearly interpolated for each day of observation using the inclinations for the July 1st, 1777, $\varepsilon_1=23°27'46.9"$, and for October 1st, 1777, $\varepsilon_2=23°27'47.3"$, but Bošković used the inclina-tion $\varepsilon=23°28'$ rounded to whole minutes. For the period of observation (September 12th to 19th, 1777) the inclination correction for September 12th, 1777 is $\Delta\varepsilon_{12}=0.4"\cdot74/92=0.32"$ and for September 19th, 1777, it is $\Delta\varepsilon_{19}=0.4"\cdot81/92=0.35"$. The correction is not significant, but we could use for the period of observation September 12th to 19th, 1777 corrected to $\varepsilon_{corr}=\varepsilon_1+\Delta\varepsilon=23°27'46.9"+0.3"=23°27'47.1"$, instead of $\varepsilon=23°28'$, which Bošković used.

**Table 40:** The observation data control: the 1st line: date; the 2nd and the 3rd lines: the observation time differences in seconds (") between observations of the set $\Delta_1$=time (tache)–time (1 bord) and $\Delta_2$=time (2 bord)–time (tache); the 4th line: the observation time differences between the observation sets $\Delta_3$=time(2 bord)–time(1 bord), the 5th line: $\Delta_4$ between neighbouring series, and $\Delta_5$ duration of daily observation, and calculated $B=t_t-(t_1+t_2)/2$ (**Boscovich, 1785 and present work**).

|   | 1 | 2 | 3 | 4 | 5 | |
|---|---|---|---|---|---|---|
| Date: | 12th | Sept. | 1777. | | | |
| $\Delta_1$= | 106 | 107 | 108 | 107 | 108 | |
| $\Delta_2$= | 21 | 21 | 20 | 21 | 81 | |
| $\Delta_4$= | **326** | **102** | **107** | **328** | **1563** | =$\Delta_5$ |
| $\Delta_3$= | 127 | 128 | 128 | 128 | **189** | 26m03s |
| B= | 42.5 | 43.0 | 44.0 | 43.0 | 13.5 | |
|   | 13th | Sept. | 1777. | | | |
| $\Delta_1$= | 97 | 97 | 97 | 98 | 98 | |
| $\Delta_2$= | 30 | 31 | 31 | 30 | 32 | |
| $\Delta_4$= | **33** | **116** | **37** | **330** | **1157** | =$\Delta_5$ |
| $\Delta_3$= | 127 | 128 | 128 | 128 | **130** | 19m17s |
| B= | 33.5 | 33.0 | 33.0 | 34.0 | 33.0 | |
|   | 15th | Sept. | 1777. | | | |
| $\Delta_1$= | 75 | 75 | 75 | 76 | -524 | |
| $\Delta_2$= | 53 | 52 | 53 | 52 | 52 | |
| $\Delta_4$= | **318** | **90** | **55** | **680** | **1182** | =$\Delta_5$ |
| $\Delta_3$= | 128 | 127 | 128 | 128 | **-472** | 19m42s |
| B= | 11.0 | 11.5 | 11.0 | 12.0 | -288.0 | |
|   | 16th | Sept. | 1777. | | | |
| $\Delta_1$= | 64 | 64 | 63 | 64 | 65 | |
| $\Delta_2$= | 64 | 64 | 65 | 64 | 64 | |
| $\Delta_4$= | **46** | **45** | **49** | **36** | **817** | =$\Delta_5$ |
| $\Delta_3$= | 128 | 128 | 128 | 128 | **129** | 13m37s |
| B= | 0.0 | 0.0 | -1.0 | 0.0 | 0.5 | |
|   | 17th | Sept. | 1777 | | | |
| $\Delta_1$= | 53 | 53 | 54 | 54 | 55 | |
| $\Delta_2$= | 74 | 74 | 73 | 74 | 73 | |
| $\Delta_4$= | **256** | **102** | **62** | **103** | **1160** | =$\Delta_5$ |
| $\Delta_3$= | 127 | 127 | 127 | 128 | **128** | 19m20s |
| B= | -10.5 | -10.5 | -9.5 | -10.0 | -9.0 | |
|   | 19th | Sept. | 1777 | | | |
| $\Delta_1$= | 37 | 36 | 35 | 36 | **36** | |
| $\Delta_2$= | 91 | 92 | 93 | 92 | **92** | |
| $\Delta_4$= | 40 | 149 | -12 | 48 | **865** | =$\Delta_5$ |
| $\Delta_3$= | 128 | 128 | 128 | 128 | 128 | 14m25s |
| B= | -27.0 | -28.0 | -29.0 | -28.0 | -28.0 | |

**Table 41:** Differences of the mean solar time *T.M.* of the sunspot, and the positions *lon.t* and *lat.B.t*: the original Bošković and the reproduction in present work.

| ΔT.M. | Δlon.t | Δlat.B.t |
|---|---|---|
| minutes | ' | ' |
| -3 | -2 | 0 |
| -3 | -2 | 0 |
| -1 | -1 | 0 |
| 3 | -5 | 2 |
| 0 | 1 | 1 |
| -1 | 6 | 1 |

### 4.6. Positions and mean solar time of the sunspot

The original positions *lon.t* and *lat.B.t* and mean solar time *T.M.* of the sunspot *Tab. II.* (see **Table 9**) and the reproduction in present work *Tab. II.* (see **Table 23**) have differences, presented in **Table 41**. The differences are not significant: –3m<ΔT.M.<3m, –5'<Δlon.t<6' and 0'<Δlat.B.t<2'. The periods T' and T'' derived from T.M. in Table 36 are almost the same. For lon.t and lat.B.t, we have not derived results yet, but the differences are not substantial, so we suggest further research. In this present work, we determined the positions using the corrected input data discussed in *4.2. Observation input data control*. For the reproduction of the results in this present work, we used Bošković's original results in *Tab. II.* (see **Table 9**). That way, we can compare the solar rotation elements in Bošković's example and in the present work.

## 5. Conclusions

The most time-consuming part of this research involves discovering "the calculation chains" for each computational process. In the beginning, many elements of the chains were missed. Later, the gaps were identified and filled, and now we have the whole chains for every part of the calculation. The most challenging part of the research was discovering the parts where the original formulas were missing. We reconstructed these formulas using Bošković's results integrated in spreadsheets for calculations. The results are presented here and later critically discussed.

Bošković determined the solar rotation elements using his own observations of one sunspot over a period of six days in September 1777. He determined the mean solar time *T.M.*, and six positons of the sunspot, the longitude and the latitude, its ecliptic coordinates *lon.t* and *lat.B.t* in the *Tab. II.* (see **Table 9**).

We reproduced the solar equator inclination *i*, the longitude of the node $\Omega$, and the period of the solar rotation *T* with Bošković's original formulas. In the present work, the results for the one sunspot observed over a period of six days are given. We successfully reproduced the whole original work (**Boscovich, 1785, pages 166-169**) resulting in very similar results in this present work.

Ruđer Bošković determined the mean solar time of *T.M.* and the geocentric positions of one sunspot, and then the ecliptic coordinates based on observations of the trajectory on the solar disk over 6 days in *Tab. I.* and *Tab. II.* Based on the mean solar time and ecliptic coordinates of the sunspot trajectory in six days of observation (see **Table 9**), Bošković determined the elements of solar rotation with his own methods: longitude of the node $\Omega$, inclination of the solar equator towards the ecliptic *i*, and the period of solar rotation sidereal T' and synodic T''. Bošković determined the longitude of the node $\Omega$ on the basis of two methods: 1. the method using two positions of the same sunspot (*2.1. The method for $\Omega$*) with 6, 8 and 10 pairs: $\Omega_6=70°21'$ (6 pairs), $\Omega_8=71°32'$ (8 pairs), $\Omega_{10}=73°09'$ (10 pairs), and 2. the method based on three positions of one sunspot (*2.4. The method for i and $\Omega$*) $\Omega_{136}=74°03'$ (positions 1, 3 and 6). The inclination of the ecliptic *i* was determined by: 1. the method based on two positions of the same sunspot and the known longitude of the node (*2.2. The method for i*) based on five pairs $i=7°44'$, and 2. the method based on three positions of one sunspot (*2.4. The method for i and $\Omega$*) $i=6°49'$. The rotation period was determined by method *2.3. The method for the period* from 6 pairs of spots sidereal T'=26.77 days and synodic T''=28.89 days (*3.3. Bošković's results*).

We reproduced the same example in the same way with Bošković's methods: mean solar time *T.M.* and the geocentric coordinates of the sunspot and then the ecliptic coordinates of the sunspot. We reproduced the elements of solar rotation with the original Bošković mean solar times and ecliptic coordinates so that we could compare the numerical values in Bošković's example, since our mean solar time and ecliptic sunspot positions are slightly different from Bošković's results (see **Table 41**). Using Bošković's methods with the same mean solar times and ecliptic coordinates of one and the same sunspot, we determined the longitude of the node $\Omega$ on the basis of two methods: 1. method using two positions of the same sunspot (*2.1. The method for $\Omega$*) with 6, 8 and 10 pairs: $\Omega_6=70°21'$ (6 pairs), $\Omega_8=71°32'$ (8 pairs), $\Omega_{10}=73°12'$ (10 pairs), and 2. the method for three positions of one sunspot (*2.4. The method for i and $\Omega$*) $\Omega_{136}=74°03'$ (positions 1, 3 and 6). The inclination of the ecliptic *i* was determined by: 1. the method based on two positions of the same sunspot and the known longitude of the node (*2.2. The method for i*) based on five pairs $i=7°45'$, and 2. the method based on three positions of one sunspot (*2.4. The method for i and $\Omega$*) $i=6°48'$. The rotation period was determined by method *2.3. The method for the period* of 6 pairs of sunspots sidereal T'=26.77 days and synodic T''=28.89 days. Periods of solar rotation were also determined based on our mean solar times *T.M.* (see **Table 34**) and determined almost identical values of T'=26.76 days and synodic T''=28.87 days of the period of solar rotation (*3.4. Present work results*).

The angular values Bošković (**Boscovich, 1785**) presented in the so-called Zodiac signs ($1^s=30°$), degrees (°) and minutes (') the angle notation that is not usual today, for example $lon.t=10^s11°42'=(10·30°+11°)42'=311°42'$ in *Tab. II.* (see **Table 9**, column two, *lon.t* for the first observation). One sign is actually the width of one sign of Zodiac, there are twelve of them in ecliptic of 360°, so 360° divided by 12 is 30°. Then this terminology was usual as we can read at the end of №114, where he calculated longitude of the node by adding three signs to D, one sign is 30° ($1^s=30°$) as we mentioned before (**Boscovich, 1785, №114, pages 136-137**).

Bošković corrected the time *T.M.* using the astronomical almanac *Connoissance des temps* (**URL 1**). We determined *T.M.* using a correction for the mean solar time which resulted in *T.M.* values that were different from those that Bošković determined. We assume that Bošković used the wrong table from the astronomical almanac (**URL 1**), but we do not have the final confirmation for this yet.

In this work, we reproduced one *T.M.* for five series of observations as an arithmetic mean of the initial time of the first and the final time of the fifth series for each day of observation. In fact, Bošković observed the sunspot five times each day. We could determine five sunspot positions and *T.M.* for each day of observation. The presumption is that Bošković used arithmetic means for *A*, *B*, the sunspot position and *T.M.*, because the procedures of his methods are relatively complex for determination with logarithmic tables. We repeated these procedures much more easily with modern computers.

This research has an application in modern astronomy for the transformation of solar rotation elements. Some telescopes operate in *Alt-Az* (*Altitude* and *Azimuth*) or *Ra-Dec* coordinates, (*Right ascension* α and *Declination* δ) not in the solar ones. Complex solar motion in coordinates *Ra-Dec/Alt-Az* includes: 1. the Sun is moving in the sky (Earth rotation + revolution), 2. the Sun is rotating differentially. 3. solar features also have proper motions. The application is for the transformation of solar coordinates in *Ra-Dec* (α, δ) and vice versa, for e.g., ALMA Solar Ephemeris Tool (Skokić and Brajša, 2019). ALMA solar images come in *Ra-Dec* system – they need to be transformed into the solar coordinates. Modern astronomy allows for better resolution and greater astrometric precision which means precise (*i*, $\Omega$). For exam-ple, errors in (*i*, Ω) give false / artificial shifts, e.g. meridional motions. Finally, it is also important for combining observations from different places, for example the Earth, satellites, etc. The present work opened many questions and widens the horizon of Bošković's thinking in his time.

Further steps for this research topic of Ruđer Bošković are:

1. in this paper, we reproduced Bošković's results *Tab. III.* to *Tab. XII.* (see **Table 24** to **Table 33**) using the time and positions of the sunspot determined by Ruđer Bošković (*T.M.* and ecliptic coordinates *lon.B.t* and *lat.t*) in *Tab. II.* (see **Table 9**). In this paper, we calculated the time *T.M.* and the ecliptic coordinates *lon.B.t* and *lat.t* in *Tab. II.* (see **Table 23**), which are slightly different from Bošković's in *Tab. II.* (see **Table 9**). Using the data *Tab. II.* (see **Table 23**) should be determined by *Tab. III.* to *Tab. XII*;
2. in September 1777, Bošković observed four sunspots presented in *Appendice* (**Boscovich 1785, Appendice, pages 170-178**). We can determine all sunspot positions with *T.M.* for determination of the solar rotation elements;
3. the original formulas can be streamlined into more convenient ones for modern computers using contemporary information technology. The streamlined formulas can be put in a convenient programming language;
4. the 1777 observation could be put into modern formulas for determination of the solar rotation elements and then compare them with Bošković's results and the results in this present work.

## Acknowledgements

This work has been supported by the Croatian Science Foundation under the project 7549 "Millimeter and submillimeter observations of the solar chromosphere with ALMA". It has also received funding from the Horizon 2020 project SOLARNET (824135, 2019–2022).

## SAŽETAK

### Određivanje elemenata Sunčeve rotacije *i*, Ω i perioda opažanjima Sunčevih pjega Ruđera Boškovića 1777. godine

U rujnu 1777. godine Ruđer Bošković šest je dana opažao Sunčeve pjege. Na osnovi tih mjerenja vlastitim je metodama izračunao elemente Sunčeve rotacije, longitudu čvora, inklinaciju Sunčeva ekvatora i period. Opis metoda, način opažanja i detaljne upute za računanje objavio je u drugome poglavlju petoga dijela *Opera* 1785. godine. U ovome radu objavljeni su originalni Boškovićevi izračuni i ponovljena su računanja njegovim postupcima. Analizom ulaznih veličina, postupaka i rezultata diskutirane su ulazne veličine, pronađene pogreške i rezultati računanja. Reprodukcija Boškovićevih izračuna uspješno je ponovila postupke i dobila vrlo slične rezultate. Zaključkom su predložena povezivanja Boškovićevih istraživanja s modernom astronomijom.

**Ključne riječi:**
Ruđer Bošković, elementi Sunčeve rotacije, opažanja Sunčevih pjega

## Author's contribution

**Mirko Husak** (MSc, PhD student) performed calculations and reproductions of the results with the discussion. **Roman Brajša** (scientific adviser with tenure) and **Dragan Špoljarić** (full professor) contributed to the interpretation and presentation of the results. **Roman Brajša** (scientific adviser with tenure) contributed to the analysis of solar rotation and its relationship to modern applications. **Dragan Špoljarić** (full professor) made substantial contributions in time issues and transformation of the coordinate systems.